\documentclass[aps,prb,showpacs,twocolumn,floatfix]{revtex4}
\usepackage{graphicx}

\newcommand{\GVB}{$G(V,B)$}
\newcommand{\micro}[1]{$\mu\text{#1}$}
\newcommand{\fig}[1]{Fig.~\ref{#1}}
\newcommand{\figs}[1]{Figs.~\ref{#1}}
\newcommand{\eq}[1]{Eq.~(\ref{#1})}

\begin{document}

\date{Draft updated on \today}

\author{Yaroslav Tserkovnyak}
\author{Bertrand I. Halperin}
\affiliation{Lyman Laboratory of Physics, Harvard University,
Cambridge, Massachusetts 02138}
\author{Ophir M. Auslaender}
\author{Amir Yacoby}
\affiliation{Dept. of Condensed Matter Physics, Weizmann
Institute of Science, Rehovot 76100, Israel}

\title{Interference and zero-bias anomaly in tunneling between
Luttinger-liquid wires}

\begin{abstract}
We present theoretical calculations and experimental
measurements which reveal the Luttinger-liquid (LL) nature of
elementary excitations in a system consisting of two quantum
wires connected by a long narrow tunnel junction at the edge of
a GaAs/AlGaAs bilayer heterostructure. The boundaries of the
wires are important and lead to a characteristic interference
pattern in measurements on short junctions. We show that the
experimentally observed modulation of the conductance
oscillation amplitude as a function of the voltage bias can
be accounted for by spin-charge separation of the
elementary excitations in the interacting wires. Furthermore,
boundaries affect the LL exponents of the voltage and
temperature dependence of the tunneling conductance at low
energies. We show that the measured temperature dependence of
the conductance zero-bias dip as well as the voltage
modulation of the conductance oscillation pattern can be used
to extract the electron interaction parameters in the wires.
\end{abstract}


\maketitle

\section{Introduction}

Quasi-one-dimensional (1D) structures with gapless electronic excitations,
such as carbon nanotubes, quantum Hall edge states, and confined states at
the edge of a quantum well heterostructure (i.e.,
quantum wires), possess unique properties which cannot be
described by Landau's Fermi-liquid theory. Even small
electron-electron interactions in a 1D confinement
make inadequate the picture based on the
existence of long-lived fermionic quasiparticles which can be mapped
onto single-particle states in a free-electron gas. A powerful framework
for understanding universal properties of 1D electron systems was put forward
by the formulation of Luttinger-liquid (LL) theory.\cite{Haldane:jpc81}
(For a review see Ref.~\onlinecite{Voit:rpp94}.)
The spectral density, $\mathcal{A}(k,\omega)$,
of the one-electron Green function in a Luttinger liquid
is fundamentally different from that of a Fermi liquid:
While the latter has one quasiparticle peak,
the former has two singular peaks
corresponding to the charge- and spin-density excitation
modes.\cite{Meden:prb92,Voit:prb93}

Tunnel-coupled quantum wires of high quality
created at a cleaved edge of GaAs/AlGaAs double-quantum-well
heterostructures appear to be an exceptional tool for probing
spectral characteristics of a 1D
system.\cite{Auslaender:sc02,Carpentier:prb02,Zulicke:prb02}
It is achieved\cite{Auslaender:sc02} by measuring the differential
conductance \GVB\ as a function of the voltage bias between the wires, $V$, and
magnetic field oriented perpendicular to the plane
of the cleaved edge, $B$, allowing for simultaneous control of
the energy and momentum of the tunneling electrons.
In a recent article\cite{Tserkovnyak:prl02} we demonstrated that the
picture of noninteracting electrons can be used with great
success to explain some of the most pronounced features of
the conductance interference pattern
arising from the finite size of the tunneling region.
Taking electron-electron interactions
into account was shown to explain experimentally observed long-period
oscillation modulations in the $V$ direction, which can be understood
as a moir\'{e} pattern arising from spin-charge separation of
electronic excitations.
In this paper we use LL formalism to further investigate an
interplay between electron correlations and the finite length of
the tunnel junction, which allows us to understand peculiarities of the
oscillations and the zero-bias anomaly in the measured tunneling conductance \GVB.

\section{Experimental method}

In this section we describe the means by which we measure the
tunneling conductance through a single isolated junction
between two parallel wires.

\subsection{Fabrication of the samples}

\begin{figure}
\includegraphics[width=3.375in,clip=]{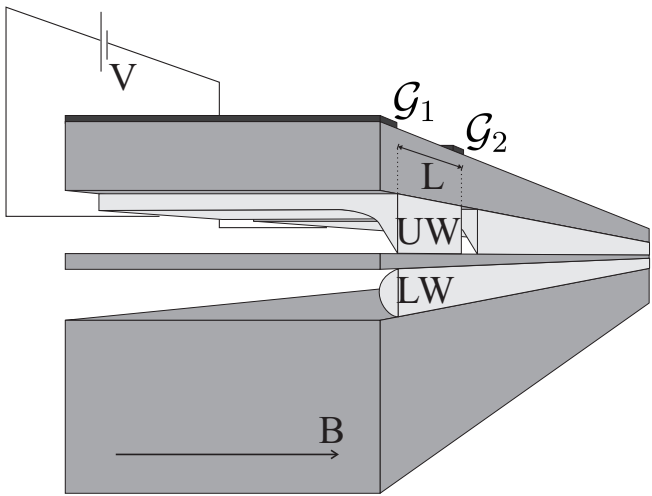}
\caption{\label{circuit}Illustration of the sample and the
contacting scheme. The sample is fabricated using the CEO
method. The parallel 1D wires
span along the whole cleaved edge (right facet in the
schematic). The upper wire (UW) overlaps the 2DEG, while the lower
wire (LW) is separated from them by a thin AlGaAs barrier
(AlGaAs is dark gray in the schematic, electron liquids are light gray).
Contacts to the wires are made through the 2DEG. Several tungsten top
gates can be biased to deplete the electrons under them: We
show only $\mathcal{G}_1$, here biased to deplete the 2DEG and both
wires, and $\mathcal{G}_2$, here biased to deplete only the 2DEG and
the upper wire. The magnetic field $B$ is perpendicular to
the plane defined by the wires. The depicted configuration
allows the study of the conductance of a tunnel junction
between a section of length $L$ of the upper wire and a
semi-infinite lower wire.}
\end{figure}

The two parallel 1D wires are fabricated by cleaved-edge
overgrowth (CEO), see \fig{circuit} and Ref.~\onlinecite{Yacoby:ssc97}.
Initially, a GaAs/AlGaAs heterostructure with two closely situated
parallel quantum wells is grown. The upper quantum well is 20~nm
wide, the lower one is 30~nm wide and they are separated by a 6~nm
AlGaAs barrier about 300~meV high. We use a modulation doping
sequence that renders only the upper quantum well occupied by a
two-dimensional electron gas (2DEG) with a density
$n\approx2\times10^{11}~\text{cm}^{-2}$ and mobility
$\mu\approx3\times10^6~\text{cm}^2\text{V}^{-1}\text{s}^{-1}$. After cleaving
the sample in the molecular beam epitaxy growth chamber and
growing a second modulation doping sequence, two parallel quantum
wires are formed in the quantum wells along the whole edge of the
sample. Both wires are tightly confined on three sides by
atomically smooth planes and on the fourth side by the triangular
potential formed at the cleaved edge.

Spanning across the sample are several tungsten top gates of width
2~\micro{m} that lie 2~\micro{m}\ from each other (two of these
are depicted in \fig{circuit}). The differential conductance $G$
of the wires is measured through indium contacts to the 2DEG
straddling tungsten top gates. While monitoring $G$ with standard lock-in
techniques (we use an excitation of $10$~\micro{V} at $14$~Hz) at
$T=0.25$~K, we decrease the density of the electrons under the
gate by decreasing the voltage on it ($V_g$).
At $V_g=V_{\text{2D}}$, the 2DEG depletes and $G$
drops sharply, because the electrons have to scatter into the
wires in order to pass under the gate. For $V_{\text{2D}}>V_g>V_U$
the conductance drops stepwise each time a mode in the upper
wire is depleted.\cite{Yacoby:prl96} In this voltage range, the
contribution of the lower wire to $G$ is negligible because it is
separated from the upper quantum well by a tunnel barrier.
When $V_g=V_U$, the upper wire depletes and only the lower wire
can carry electrons under the gate. This last conduction channel
finally depletes at $V_L$ and $G$ is suppressed to zero.

\subsection{Measurement on an isolated tunnel junction}

The measurements are performed in the configuration depicted
in \fig{circuit}. The source is the 2DEG between two gates,
$\mathcal{G}_1$ and $\mathcal{G}_2$ in \fig{circuit}, the voltages on which
are $V_1<V_L$ and $V_L<V_2<V_U$, respectively. The upper wire
between these gates is at electrochemical equilibrium with
the source 2DEG. This side of the circuit is separated by the
tunnel junction we wish to study from the drain. The drain is
the 2DEG to the right of $\mathcal{G}_2$ (the semi-infinite 2DEG in
\fig{circuit}) and it is in equilibrium with the right,
semi-infinite, upper wire and with the whole semi-infinite
lower wire in \fig{circuit}. Thus, any voltage difference ($V$)
induced between the source and the drain drops on the
narrow tunnel junction between the gates. This configuration
gives us control over both the energy and the momentum of the
tunneling electrons, as explained below. An additional gate
lying between $\mathcal{G}_1$ and $\mathcal{G}_2$ (not shown in \fig{circuit})
allows us to deplete the 2DEG in the center of the source,
thus reducing the screening of the interactions in the wires
by the 2DEG.

The energy of the electrons tunneling between the wires is
given by $eV$, $-e$ being the electron charge. The
tunneling process occurs along the whole length $L$ of the
tunnel junction. Therefore, momentum is conserved to within
an uncertainty of order $2\pi/L\ll k_{\text{F}}$,
where $k_{\text{F}}$ is a typical Fermi wave vector in the wires.
We can shift the momentum of the tunneling electrons with a magnetic
field ($B$) perpendicular to the plane defined by the wires. The
value of the shift is $\hbar q_B=eBd$, where $d$ is the
center-to-center distance between the wires.

\section{Description of the experimental results}
\label{exp}

In the experiment we measure the nonlinear differential
tunneling conductance \GVB\ through a junction between two
parallel wires. The sample we report here contains four top
gates allowing us to vary the length of the junction $L$ by
choosing different combinations of gates. We have studied in
detail junctions with $L=2,~4,~6,~10$~\micro{m} as well as
symmetric junctions ($L=\infty$). The results presented here
are from junctions with $L=2,~6,~10$~\micro{m}.
Many of the effects that we measure rely on the smallness of $1/L$,
while others (which we address here in detail) are present only when
$L$ is finite.

\subsection{Dispersions of elementary excitations in the wires}
\label{disp}

\begin{figure}
\includegraphics[width=3.375in,clip=]{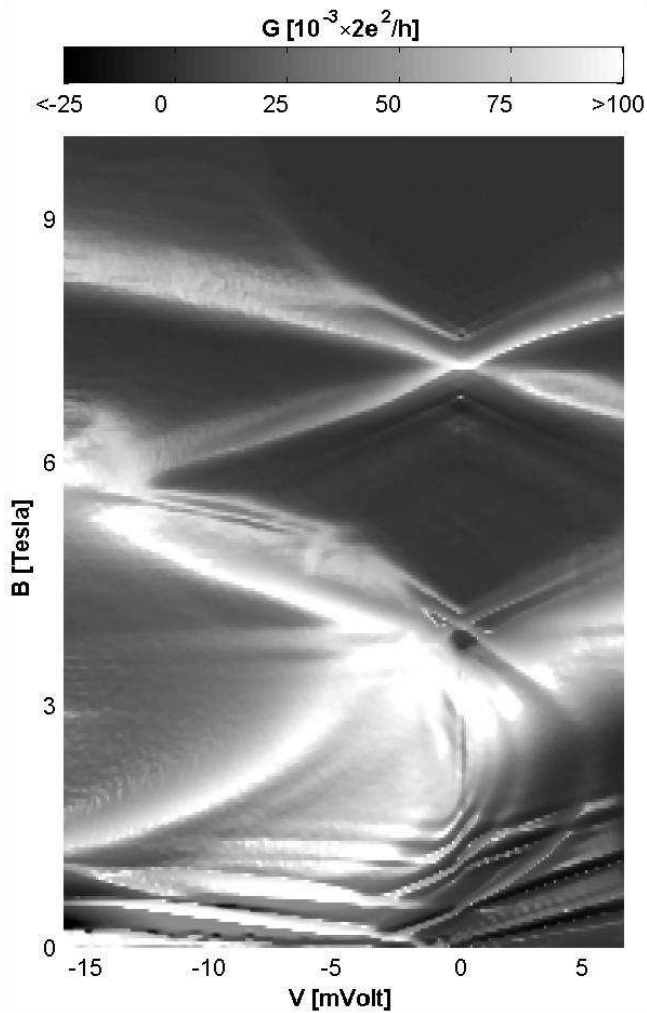}
\caption{\label{GVB}Plot of \GVB\ for a 10~\micro{m}\
junction. Higher values of the conductance are depicted in lighter shades:
The top bar gives the key.}
\end{figure}

By mapping out \GVB\ we determine the dispersion curves of
the wires.\cite{Auslaender:sc02} These are
given by the curves that are traced by the main peaks
as seen in \fig{GVB}. We can understand their gross features
employing a noninteracting electron picture:\cite{Auslaender:sc02}
The peaks result from tunneling
between a Fermi point in one wire and a mode in the other
wire. Since each occupied mode has two Fermi points, two
copies of the dispersion show up in the \GVB\ scan. All in
all, for each pair of occupied modes in the two wires we expect to observe
four dispersions, because there are four Fermi points
involved: $\pm k^i_{\text{F}u}$ and $\pm k^j_{\text{F}l}$.
(Indices $i$ and $j$ label various modes in the wires, $u$
and $l$--the upper and the lower wires.)
In reality, we observe only some of the transitions:
For example, by carefully studying \fig{GVB}
one can distinguish dispersions of three modes in the upper wire and
five in the lower one, but only the following
transitions seem to have a sizable signal:
$\left|u_1\right>\leftrightarrow\left|l_1\right>$,
$\left|u_3\right>\leftrightarrow\left|l_2\right>$, and
$\left|u_2\right>\leftrightarrow\left|l_{3,4,5}\right>$,
where the order in the list is of decreasing $B_2^{i,j}$ (see
below); $\left|u_3\right>$ is the 2DEG occupying the upper
quantum well.\cite{Auslaender:sc02}
Such selection rules are related to the shape of the
wave functions in the direction perpendicular to the cleaved edge.
In identical wires, one would expect only the transitions
$\left|u_n\right>\leftrightarrow\left|l_n\right>$ to show up,
due to the orthogonality between different modes.
In dissimilar wires, the selection rules are different and
less strict so other transitions are observed.

The dispersions allow us to extract the densities of
electrons in each mode, $n^i_{u(l)}=(2/\pi)k_{\text{F}u(l)}^i$, as follows.
Tunneling amongst each pair of
occupied modes is enhanced near $V=0$ at two values of $B>0$,
where the two curves in \GVB\ cross. In the first (in the
following referred to as the \textquotedblleft lower crossing
point\textquotedblright), which occurs at $B_1^{i,j}$, the
direction in which the electrons propagate is conserved in
the tunneling process. In the second (referred to as the
\textquotedblleft upper crossing point\textquotedblright),
the Lorentz force exerted by $B_2^{i,j}$ exactly compensates
for the momentum mismatch between oppositely moving electrons
and the direction of propagation of the tunneling electrons
reverses. In wires with vanishing cross section, these
crossing points occur at
\begin{equation}\label{B12}
\left|B_{1(2)}^{i,j}\right|=
\frac{\hbar}{ed}\left|k_{\text{F}u}^i\mp
k_{\text{F}l}^j\right|\,.
\end{equation}
In principle \eq{B12} can be used to extract the densities of
the modes, regardless of electron-electron interactions
in the wires\cite{Carpentier:prb02}
or mesoscopic charging\cite{Boese:prb01}
that can merely smear them at a finite voltage bias. In
realistic wires that have a finite cross section, finding the
densities is hampered by the weak magnetic field dependence
that they acquire. This difficulty is overcome by a simple
fitting procedure that we have developed: We assume that all
the modes in a wire have the same field dependence, a
reasonable assumption for our tight-confining potential
in the growth direction of the quantum wells. We
then guess the $B=0$ occupations of the modes in each wire,
$n_u^i(0)$ and $n_l^j(0)$, and calculate their field dependences.
If the resulting dispersions do not cross at $B_{1(2)}^{i,j}$,
we adjust $n_u^i(0)$ and $n_l^j(0)$ and repeat the procedure.
This is done iteratively for all the crossing points that we
see, because changing the occupation of one mode affects the
field dependence of all the other occupations in a wire. The
dispersion that we use is that of noninteracting electrons
in a finite quantum well, in the presence of an in-plane magnetic field.
Such a dispersion depends only on the width and depth of the well
and on the band mass of electrons in GaAs.

In every case we have studied, we see clear deviations of the measured
dispersions from the calculated noninteracting ones at a finite bias. In
particular, we find that the velocities of some excitations
are enhanced relative to the Fermi velocities
$v_{\text{F}u(l)}$. The former are given by
\begin{equation}
v_p=\frac{1}{d}\left.\frac{\partial V}{\partial B}
\right|_{B_{1(2)}^{i,j}} \label{vp}
\end{equation}
(along the observed main peaks),
while the latter can be obtained by the calculated slope of the
(noninteracting) dispersions at the Fermi points. This velocity enhancement
is thought to correspond to the charge-density modes and
can be accounted for by electron-electron interactions in the
wires.\cite{Carpentier:prb02,Auslaender:sc02}

The ability to determine the dispersion relations relies on
the high quality of the junctions to sustain momentum-conserving
tunneling. Momentum relaxation ensues as soon as
invariance to translations is broken. The most obvious
mechanism by which this happens is the finiteness of $L$. We
find that we indeed observe its effects. The second mechanism is the
disorder inherent to all semiconductor devices, some effects
of which seem to also be observed.

\subsection{Oscillations}
\label{osc}

\begin{figure}
\includegraphics[width=3.375in,clip=]{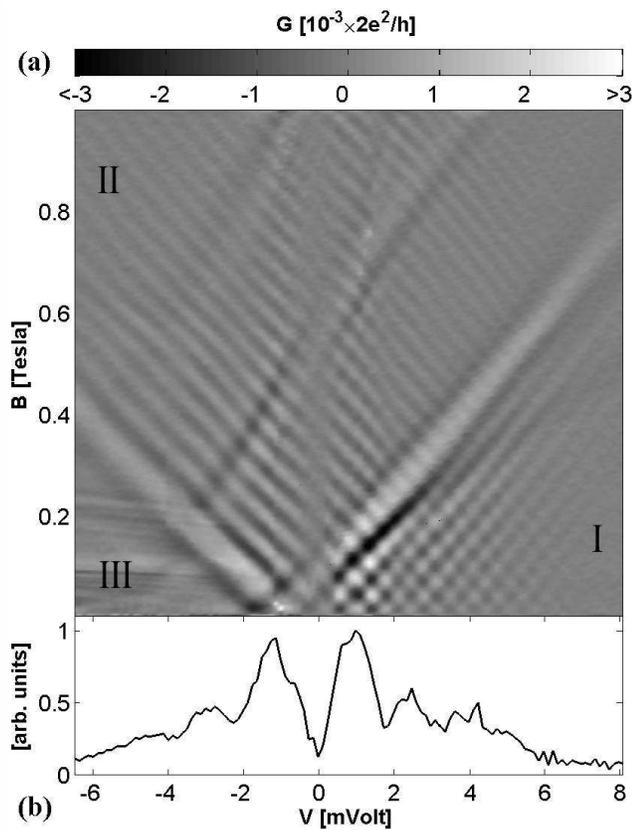}
\caption{\label{osc24} Nonlinear conductance oscillations at
low field from a 2~\micro{m} junction. (a) shows the
oscillations as a function of both $B$ and $V$. (A smoothed
background has been subtracted to emphasize the
oscillations.) The brightest (and darkest) lines, corresponding to
tunneling between the lowest modes, break the $V$-$B$ plain into
regions I, II, and III. Additional positively-sloped bright and dark lines
in II arise
from other 1D channels in the wires and are disregarded in
our theoretical analysis. Also present is a slow modulation
of the strength of the oscillations along the abscissa.
(b) Absolute value of the peak of the Fourier transform
of $S^{1-1/\beta}G\left(V,S^{1+1/\beta}\right)$ at a fixed $V$ in region
$II$ as a function of $V$.
(See Sec.~\ref{asb} for definition of $S$, $\beta$ and other details.)
Its slow modulation as a function of $V$ is easily discerned.}
\end{figure}

\begin{figure}
\includegraphics[width=3.375in,clip=]{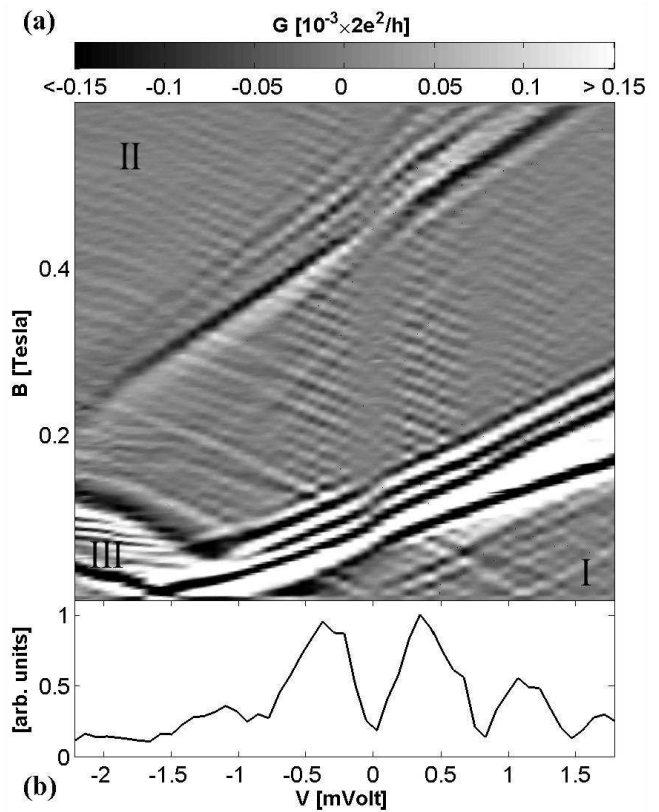}
\caption{\label{osc26}Same as \fig{osc24} but for a
6~\micro{m} junction. Note that the oscillations are
approximately three times faster than in \fig{osc24}, as
expected from \eq{AB}. For this junction, there are several additional
side lobes present on the left of the principal peaks, unlike
in the case of the shorter junction in \fig{osc24}.}
\end{figure}

The most spectacular manifestation of the breaking of
translational invariance is the appearance of a regular
pattern of oscillations away from the dispersion curves.
\figs{osc24}a and \ref{osc26}a are typical examples of the
patterns that we measure at low magnetic field. In this range
of field, the lines that correspond to the dispersion curves
appear as the pronounced peaks that extend diagonally across
the figures. In addition to these we observe numerous
secondary peaks running parallel to the main dispersion curves.
These side lobes always appear to the right of
the wire dispersions, in the region that corresponds to
momentum conserving tunneling for an upper wire with a
reduced density. As a result, we see a checkerboard pattern
of oscillations in region I, a hatched pattern in region II,
and no regular pattern in region III (see \figs{osc24}a and
\ref{osc26}a for the definitions).

\begin{figure}
\includegraphics[width=3.375in,clip=]{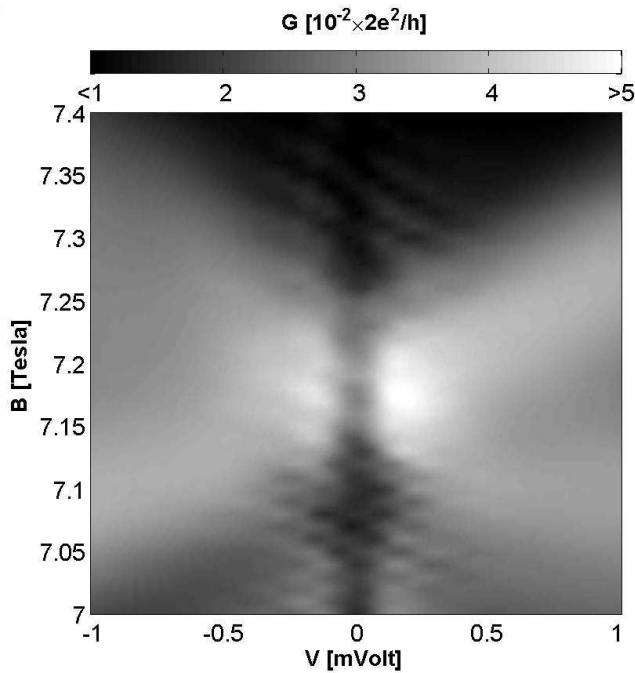}
\caption{\GVB\ near the upper crossing point for a
6~\micro{m}\ junction. In this measurement, a central
2~\micro{m}\ gate midway between $\mathcal{G}_1$ and $\mathcal{G}_2$ is biased to
deplete all upper wire modes except the lowest one. One can
see a pattern of oscillations around the dispersion peaks.}
\label{cross}
\end{figure}

The interference pattern also appears near the upper crossing
point at high magnetic field. A typical example is shown in \fig{cross}.

The frequency of the oscillations depends on $L$. When $L$ is
increased from 2~\micro{m}, \fig{osc24}, to
6~\micro{m}, \fig{osc26}, the frequency in bias
($\Delta V$) and in field ($\Delta B$) increases by about a factor
of three. The period is approximately related to the length
of the junction through the formula
\begin{equation}
\Delta V L/v_{\text{F}}=\Delta BLd=\phi_0\,, \label{AB}
\end{equation}
where $\phi_0=2\pi\hbar/e$ is the quantum of flux.

A close examination of the low-field oscillations reveals an
interesting behavior of their envelope. Notable is the
suppression of \GVB\ near $V=0$ which is independent of
field. Also visible are faint vertical gray stripes, where the
amplitude of the oscillations in the $B$ direction is reduced.
The modulation of the oscillation amplitude, as a function of $V$,
is shown in panels (b) of figures \ref{osc24} and \ref{osc26}.
The oscillatory part of $G$ thus depends on $V$ on two major scales:
The faster scale (0.5~mV for $L=2$~\micro{m}) corresponds to the
oscillations described by \eq{AB}. The slower scale (2~mV for
$L=2$~\micro{m}) governs the distance between the stripes
of suppressed \GVB\ parallel to the field axis, including the
zero-bias suppression. Like the fast scale, the slow scale is
roughly inversely proportional to the lithographic length of the
tunneling region.

\subsection{A dip in the tunneling conductance}
\label{dtc}

\begin{figure}[ptb]
\includegraphics[width=3.375in,clip=]{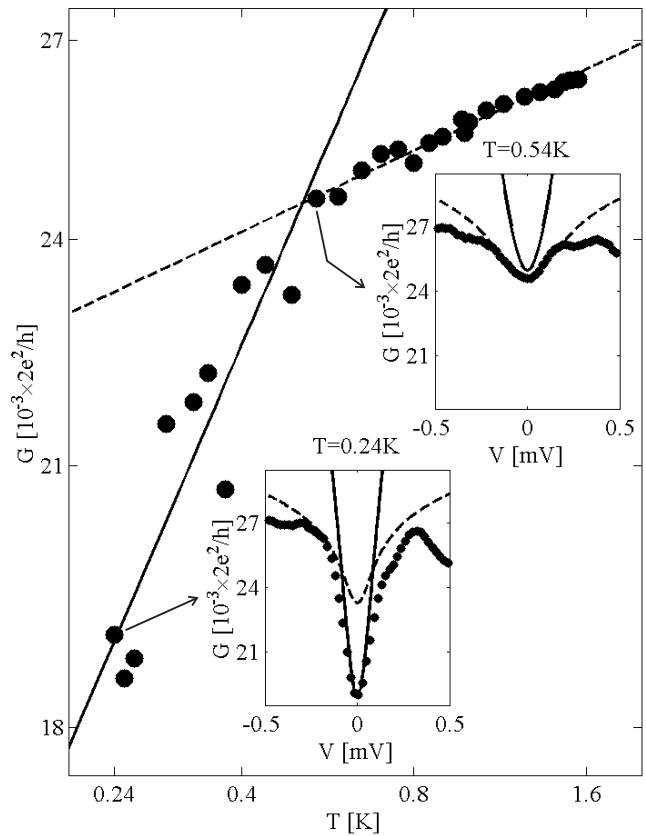}
\caption{Zero-voltage dip of the tunneling conductance $G$ as
a function of temperature on a log-log scale. The circles
show measurements on a 6~\micro{m} junction at $B=2.5$~T, the
lines are a fit using $G\propto T^\alpha$ for $V=0$. The dashed
line is the result for $\alpha=\alpha_{\text{bulk}}(g_l)=0.07$
while the solid line is the result for
$\alpha=\alpha_{\text{end}}(g_l)=0.35$, with $g_l=0.59$ (and $g_u=1$) in
Eqs.~(\ref{ab}) and (\ref{ae}), respectively;
see Sec.~\ref{zb} for a discussion.
Insets: $G(V)$ for $T=0.24$~K and $T=0.54$~K (the
temperature dependence was generated from the $V=0$ point of such scans). The
curves were calculated with \eq{GVT} and using
the above value of $g_l$ extracted
from the fit of the temperature dependence of
the dip. [We obtained $F_\alpha(x)$ by convoluting the derivative of the Fermi distribution in the 2D leads, $[1/(4k_BT)]\text{sech}^2[eV/(2k_BT)]$, with the finite-temperature tunneling density of states in the lower wire, see Eq.~(5) in Ref.~\onlinecite{Bockrath:nat99}.] The dashed lines correspond to the
$\alpha_{\text{bulk}}$ value of the exponent while the solid lines to
$\alpha_{\text{end}}$.} \label{zbd}
\end{figure}

Prominent in all scans that have high enough resolution in
$V$ is a strong suppression of the conductance near $V=0$ at
all magnetic fields. The width of this conductance dip is of
order of 0.1~mV, see Figs.~\ref{GVB} and \ref{cross}. The
size of the dip is very sensitive to temperature, as depicted
in \fig{zbd}, and it exists for $T\lesssim1.0$~K.

\section{Theory and discussion}

The 1D modes in the upper quantum well are coupled to the 2DEG
\textit{via} an elastic 1D-2D scattering which ensures a good
electronic transfer between the extended and confined states of
the well.\cite{Picciotto:prl00} In addition to
tunneling between the confined states in the wires, if the
extended states have an appreciable weight at the edge, there will
be a direct transition from the 2DEG to the lower wire. With this
in mind, we separate the total current into two contributions, one
due to tunneling between 1D bands and the other due to direct
tunneling from the 2DEG. As explained in Sec.~\ref{disp}, each of
the wires carries several 1D modes. In our analysis and comparison
with the experiment, we will only consider the transition between
the lowest 1D bands of the wires (i.e., the bands with
the largest Fermi momentum),
$\left|u_1\right>\leftrightarrow\left|l_1\right>$, and the direct
tunneling from the 2DEG,
$\left|u_3\right>\leftrightarrow\left|l_2\right>$, which
both have a strong signal, as seen in \fig{GVB}. In each wire, the
1D modes interact with each other, but since the bands have very
different Fermi velocities, we treat them independently. This is a
reasonable approximation, as explained in Appendix~\ref{a1}.

\begin{figure}[ptb]
\includegraphics[width=3.375in,clip=]{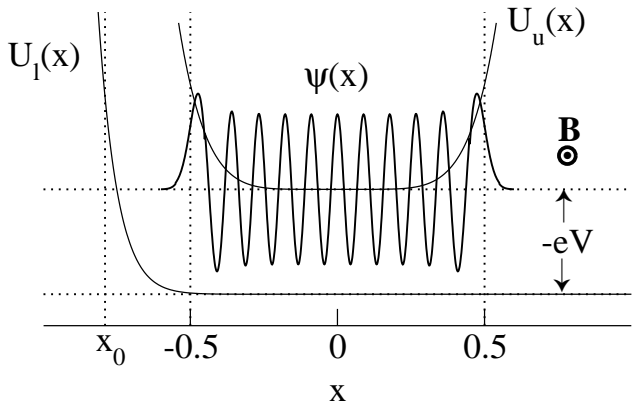}
\caption{\label{sc} Schematic picture of the theoretical model.
The upper wire is formed by a potential well $U_u(x)$
created by gates $\mathcal{G}_1$ and $\mathcal{G}_2$ (shown in \fig{circuit})
and the lower wire
is semi-infinite with the left boundary $U_l(x)$ at gate $\mathcal{G}_1$.
$\psi(x)$ is an electron wave function in the upper wire. The energy and
momentum of the tunneling electrons are governed by the voltage
bias $V$ and magnetic field $B$.}
\end{figure}

The geometry for our theoretical description is shown in
\fig{sc}. The potentials $U_u(x)$ and $U_l(x)$ are
felt by electrons in the upper and lower quantum
wires, respectively. The electrons in the upper wire are confined to a region
of finite length by potential gates at both its ends (see the
source region in \fig{circuit}). One of these gates ($\mathcal{G}_1$)
causes the electrons in the lower wire to be reflected at one end,
but the other ($\mathcal{G}_2$) allows them to pass freely under it. The
effective tunneling region is determined by the length of the
upper wire, which is approximately the region $|x| < L/2$ in
\fig{sc}. The magnetic field, $B$, gives a momentum boost
$\hbar q_B=eBd$ along the \textit{x}-axis for the electrons
tunneling from the upper to the lower wire.

First, we develop a general formalism in Sec.~\ref{for}. We
then apply it to study the conductance interference pattern
in Sec.~\ref{int} and the zero-bias anomaly regime in Sec.~\ref{zba}.

\subsection{General formalism}
\label{for}

Let us first consider transport between two 1D bands in the
wires. We use the following model Hamiltonian to study the
intermode tunneling in the system:
\begin{equation}
H=\sum_{\nu=u,l}H_0^\nu
+\sum_{\nu\nu^\prime=u,l}H_{\text{int}}^{\nu\nu^\prime}
+H_{\text{1D-2D}}+H_{\text{tun}}\,.
\end{equation}
$H_0^\nu$ is the kinetic energy of the electrons [$\nu=u~(l)$
labels the upper (lower) wire], $H_{\text{int}}^{\nu\nu}$
[$H_{\text{int}}^{ul}$] describes spin-independent
electron-electron interactions in (between) the wires,
$H_{\text{1D-2D}}$ is an effective Hamiltonian for the 1D-2D
scattering of electrons in the top quantum well, and
$H_{\text{tun}}$ is the tunneling Hamiltonian:
\begin{eqnarray}
H_0^\nu&=&v_{\text{F}\nu}\sum_s\int dx\nonumber\\
&&\times\left[\Psi^\dagger_{Rs\nu}(-i\partial_x)\Psi_{Rs\nu}-\Psi^\dagger_{Ls\nu}
(-i\partial_x)\Psi_{Ls\nu}\right]\,,\nonumber\\\\
H_{\text{int}}^{\nu\nu^\prime}&=&\frac{1}{4\pi}\sum_{ss^\prime}
\int dk\tilde{V}_{\nu\nu^\prime}(k)\left[2\rho_{Rs\nu}(k)
\rho_{Ls^\prime\nu^\prime}(-k)\right.\nonumber\\
\label{Hint}
&&+\left.\rho_{Rs\nu}(k)\rho_{Rs^\prime\nu^\prime}(-k)+\rho_{Ls\nu}(k)
\rho_{Ls^\prime\nu^\prime}(-k)\right]\,,\nonumber\\\\
\label{Htun}
H_{\text{tun}}&=&\lambda\sum_s\int dx\Psi^\dagger_{su}\Psi_{sl}
e^{-iq_Bx}+\text{H.c.},
\end{eqnarray}
where $s$ and $s^\prime$ are spin indices, $\Psi_{s\nu}$ is the
spin-$s$ electron field operator, $\Psi_{Rs\nu}$ and $\Psi_{Ls\nu}$
are the field operators for the right and left movers,
respectively,
$\Psi_{s\nu}=e^{ik_{\text{F}\nu}x}\Psi_{Rs\nu}+e^{-ik_{\text{F}\nu}x}\Psi_{Ls\nu}$,
$\rho_{Rs\nu}(k)=\int dxe^{ikx}\Psi^\dagger_{Rs\nu}\Psi_{Rs\nu}$ is the
density-fluctuation operator for the spin-$s$ right movers (and
analogously for the left movers), and
$\tilde{V}_{\nu\nu^\prime}(k)=\int dxe^{ikx}V_{\nu\nu^\prime}(x)$ is the
Fourier transform of the two-particle interaction potential
$V_{\nu\nu^\prime}(x)$.
Writing $H_{\text{int}}$ in terms of the interactions between
electrons of fixed chirality in \eq{Hint} is possible after
restricting electron correlations to small momentum transfer scattering,
e.g., if $\tilde{V}(k)\propto\exp(-r_c|k|)$ with $1/r_c\ll k_{\text{F}}$.
(By making this approximation we disregard backward and Umklapp
scattering processes, which are thought to be unimportant
in our cleaved-wire structure, see, e.g., Ref.~\onlinecite{Carpentier:prb02}.)

The 1D-2D scattering randomizes the direction of the
1D electrons in the top quantum well with a mean free path
$l_{\text{1D-2D}}\approx6$ $\mu$m.\cite{Picciotto:prl00} In
infinite wires, this weak scattering can be taken into account by
rounding the 1D electron-gas
spectral function by a Lorentzian of half width
$\Gamma=1/(2\tau_{\text{1D-2D}})$, where $\tau_{\text{1D-2D}}$ is
the 1D-2D scattering time.

If there were no interactions between the wires, i.e., $V_{ul}\equiv0$,
low-energy spin and charge excitations in each wire would
propagate with velocities $v_{s\nu}=v_{\text{F}\nu}$ and
$v_{c\nu}=v_{\text{F}\nu}/g_\nu$, respectively.
The parameters $g_\nu$ can be obtained by bosonization as
\begin{equation}
g_\nu=\left[1+\frac{2\tilde{V}_{\nu\nu}(0)}{\pi\hbar v_{\text{F}\nu}}\right]^{-1/2}<1\,,
\label{gn}
\end{equation}
in the case of repulsive interactions, $\tilde{V}_{\nu\nu}(0)>0$.
In the limit of a free-electron gas, $\tilde{V}_{\nu\nu}(0)=0$, $g_\nu=1$.

We treat tunneling between the wires to lowest order in
perturbation theory. Mesoscopic charging effects, such as discussed in, e.g.,
Ref.~\onlinecite{Boese:prb01}, are disregarded in our analysis.
The current (for electrons of each spin) is given by
\begin{equation}
I=e|\lambda|^2\int_{-\infty}^\infty dxdx^\prime
\int_{-\infty}^\infty dte^{iq_B(x-x^\prime)}e^{ieVt/\hbar}C(x,x^\prime;t)\,,
\label{Ie}
\end{equation}
where $C(x,x^\prime;t)$ is a two-point Green function
\begin{equation}
C(x,x^\prime;t)=\left\langle\left[\Psi^\dagger_l\Psi_u(x,t),
\Psi^\dagger_u\Psi_l(x^\prime,0)\right]\right\rangle\,.
\label{C}
\end{equation}
In the limit of vanishing interactions between the wires, it reduces to
\begin{eqnarray}
C(x,x^\prime;t)&=&G^>_u(x,t;x^\prime,0)G^<_l(x^\prime,0;x,t)\nonumber\\
&&-G^<_u(x,t;x^\prime,0)G^>_l(x^\prime,0;x,t)
\end{eqnarray}
expressed in terms of the one-particle correlation functions
\begin{eqnarray}
\label{ggr}
G^>_\nu(x,t;x^\prime,t^\prime)&=&-i\left\langle
\Psi_\nu(x,t)\Psi_\nu^\dagger(x^\prime,t^\prime)\right\rangle\,,\\
\label{gle}
G^<_\nu(x,t;x^\prime,t^\prime)&=&i\left\langle
\Psi_\nu^\dagger(x^\prime,t^\prime)\Psi_\nu(x,t)\right\rangle\,.
\end{eqnarray}
Note: Throughout this paper, as in the above equations,
the correlation functions are
defined for electrons with a fixed spin orientation and the spin index
is therefore omitted.

The results of this section also hold for direct 2DEG-1D
tunneling, if we define $\Psi_{su}(x,t)$ as the field operator for the
2DEG at the edge of the upper quantum well.

\subsection{Interference pattern}
\label{int}

As discussed in Sec.~\ref{osc}, the breaking of translational invariance due to the finite size of the tunneling junction can result in an oscillatory dependence of the conductance $G$ on voltage bias $V$ and magnetic field $B$. In this section we discuss in detail this behavior that arises due to interference of electrons tunneling through a finite-sized window. We show that our theoretical framework can quantitatively explain the conductance oscillations observed near the crossing points.

In the following, we mainly focus on the analysis of very distinct interference patterns measured at low magnetic fields (as in Figs.~\ref{osc24} and \ref{osc26}). In Sec.~\ref{ucp} we briefly comment on the conductance near the upper crossing point at high fields (as in \fig{cross}). It appears likely that while in the former regime the translational invariance is broken due to the finiteness of the tunneling region only, in the latter case some other mechanisms may also play a prominent role.

In the actual experiments, several 1D electron modes are occupied in
the wires. Here we consider only tunneling between modes
which have the lowest energy of transverse motion,
and hence the largest
Fermi momentum along the wire, namely $\left|u_1\right>$ and
$\left|l_1\right>$. These modes have densities that differ by only
a few percent (see Ref.~\onlinecite{Auslaender:sc02}). We thus make a simplifying
approximation $v_{\text{F}u}=v_{\text{F}l}=v_{\text{F}}$, which is
justified by the measured dispersion slopes.\cite{Auslaender:sc02}

\subsubsection{Asymmetry due to soft boundaries}
\label{asb}

In Ref.~\onlinecite{Tserkovnyak:prl02} we showed that the
observed asymmetry of the secondary oscillation peaks on the
two sides of the main dispersion curves (see
Figs.~\ref{osc24}, \ref{osc26}) can be explained within a
noninteracting electron picture and assuming a soft confining
potential $U_u(x)$ for the upper wire. Here we will employ
the model developed there to quantitatively study the form of
$U_u(x)$.

Using the phenomenological tunneling Hamiltonian (\ref{Htun}), we express
the current through the junction at zero temperature
\begin{equation}
I\propto\text{sgn}(V)\sum_m |M(n,q_B,V)|^2
\label{I1b}
\end{equation}
in terms of the tunneling matrix element
\begin{equation}
M(n,q_B,V)=\int dx\psi_n^\ast(x)e^{-iq_Bx}\varphi_{k_l}(x)
\label{Mb}
\end{equation}
between the upper-wire state $\psi_n$ and the lower-wire
state $\varphi_{k_l}$, the energy of which is lower by $eV$.
The summation in \eq{I1b} is over the integers
$[\text{sgn}(V)-1]/2<m<e|V|L/(\pi\hbar v_{\text{F}})$ denoting
the offset of the $\psi_n$ index $n = n_{\text{F}}
+\text{sgn}(V)m$ with respect to the state
$\psi_{n_{\text{F}}}$ just below the Fermi energy of the upper
wire (and linearizing the dispersions near the Fermi points,
assuming $e|V|$ is not too large).
The current (\ref{I1b}) can be expressed by a single sum
because the states of the confined upper wire are discrete,
while the states in the lower wire $\varphi_{k_l}(x)=e^{\pm
ik_lx}$ can be indexed by a continuous wave vector $k_l$. (As
in Ref.~\onlinecite{Tserkovnyak:prl02}, it is assumed that the
left boundary $U_l(x)$ of the lower wire lies outside the
tunneling region.) Since the Zeeman energy in GaAs is small,
we ignore the spin degrees of freedom. 

We argued\cite{Tserkovnyak:prl02} that for practical purposes
of understanding our  measurements, the sum in \eq{I1b} can
be replaced by an integral
\begin{equation}
I\propto\int_0^{eV}d\epsilon|M(E_{\text{F}u}+\epsilon,q_B,V)|^2
\label{I2b}
\end{equation}
labeling states in the upper wire by energy
$\epsilon$ with respect to the Fermi energy
$E_{\text{F}u}$. For the conductance obtained by
differentiating the current, this approximation will smear
out the $\delta$-functions appearing when the chemical
potential of the upper wire crosses each discrete energy
level. Physically, such smearing can be caused by 1D-2D
scattering and finite temperature. But even at low
temperatures and vanishing scattering, the result obtained by
integration [\eq{I2b}] will not be far off from that found by
summation [\eq{I1b}] as the dominant contribution to the
oscillation pattern near the lower crossing point comes from
differentiating the summand in \eq{I1b} [or correspondingly the
integrand in \eq{I2b}], as explained below.

We linearize the dispersions about the Fermi wave
vectors $k_{\text{F}\nu}$, so that $k_l$ is given by
$(k_l-k_{\text{F}l})v_{\text{F}}=(\epsilon-eV)/\hbar$. The
wave vector inside the upper wire similarly depends on
energy: $(k_u-k_{\text{F}u})v_{\text{F}}=\epsilon/\hbar$.
The matrix element squared
$|M(E_{\text{F}u}+\epsilon,q_B,V)|^2$ can then be written as
a sum of contributions due to tunneling between
right movers and between left movers,
\begin{equation}
|M(E_{\text{F}u}+\epsilon,q_B,V)|^2=|M(\kappa_+)|^2+|M(\kappa_-)|^2\,,
\label{inc}
\end{equation}
where $\kappa_\pm=k_u-k_l\pm q_B=\Delta k_{\text{F}}+eV/(\hbar v_{\text{F}})
\pm q_B$ and $\Delta
k_{\text{F}}=k_{\text{F}u}-k_{\text{F}l}$. The tunneling
matrix element
\begin{equation}
M(\kappa)=\int dxe^{i\kappa x}\psi_u(x)e^{-i k_{\text{F}u}x}
\label{m}
\end{equation}
is determined by the form of the bound-state wave function $\psi_u(x)$ at
the Fermi level of the upper wire. We wrote the right-hand side of \eq{inc}
as an incoherent sum of the contributions of the two chiralities.
This is an approximation we make by disregarding
additional interference arising due to the reflection of
electrons in the lower
wire under gate $\mathcal{G}_1$ (i.e., by the potential $U_l$ in \fig{sc}).
Taking the latter into account does not considerably affect our results.

$|M(\kappa_\pm)|^2$ do not depend on energy $\epsilon$, and
the current (\ref{I2b}) can, therefore, be written
as\cite{Tserkovnyak:prl02}
\begin{equation}
I\propto V\left[|M(\kappa_+)|^2+|M(\kappa_-)|^2\right]\,.
\label{ivm}
\end{equation}
The differential conductance $G=\partial I/\partial V$
corresponding to the current $I\propto V|M(V)|^2$ becomes
$G\propto|M(V)|^2+V\partial|M(V)|^2/\partial V$. If, for
example, the oscillatory component of $|M(V)|^2$ has the form
$\sin(\text{const}\times V)$, the amplitude of the second
term in the conductance will be $2\pi N$ times larger than
the amplitude of the first term after $N$ periods of
oscillation. The dominant contribution to the oscillatory
component of the conductance near the lower crossing point is
thus
\begin{equation}
G\propto V\frac{\partial}{\partial V}\left[|M(\kappa_+)|^2+|M(\kappa_-)|^2\right]\,.
\label{dG}
\end{equation}

\begin{figure}[ptb]
\includegraphics[angle=-90,width=3.375in,clip=]{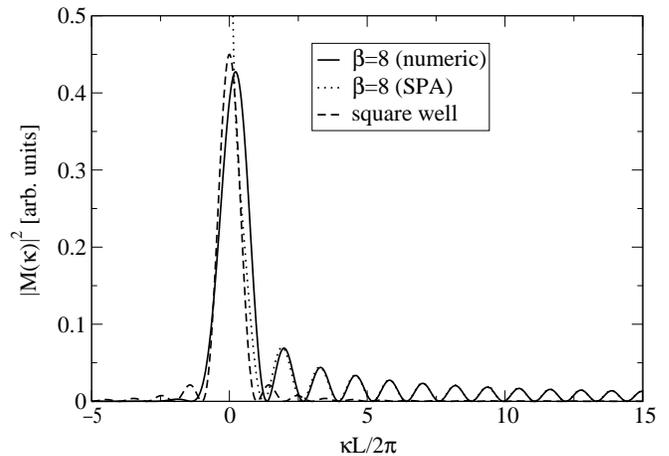}
\caption{\label{m2} $|M(\kappa)|^2$ obtained using
wave function $\psi_u$ for the 100th WKB state in the
potential well [\eq{u} with $\beta=8$] of the upper wire.
The solid line is the numerical calculation, the dotted line is the SPA
approximation [\eq{mpm}] and the dashed line shows the result for the
square-well confinement, for comparison.}
\end{figure}

If the upper wire confining potential $U_u$ is smooth enough
so that the states at the Fermi energy can be evaluated by
the WKB approximation, the form of $M(\kappa)$ [\eq{m}] can
be studied both numerically and
analytically.\cite{Tserkovnyak:prl02} In the region between
the classical turning points,
\begin{equation}
\psi_u(x)=\frac{1}{\sqrt{k_u(x)}}e^{i k_{\text{F}u}x}e^{-is(x)}\,,
\label{WKB}
\end{equation}
where $k_u(x)=k_{\text{F}u}[1-U_u(x)/E_{\text{F}u}]^{1/2}$ and
$s(x)=\int_0^xdx^\prime[k_{\text{F}u}-k_u(x^\prime)]$.
In the stationary-phase approximation (SPA),
$M(\kappa)$ is evaluated near positions $x^\pm$
($x^+>x^-$) where $k_u(x^\pm)=k_{\text{F}u}-\kappa$ and
the integrand in \eq{m} has a stationary phase. In the case
of a symmetric potential, $U_u(x)=U_u(-x)$, the SPA gives
\begin{equation}
M(\kappa)\propto\frac{\Theta(\kappa)}{\sqrt{U_u^\prime(x^+)}}
\cos\left[\kappa x^+-s(x^+)-\pi/4\right]\,,
\label{mpm}
\end{equation}
where $\Theta(\kappa)$ is the Heaviside step-function, the prime in
$U_u^\prime$ denotes the derivative. The
SPA approximation (\ref{mpm}) diverges for small values of
$\kappa$ and we have to resort to a numerical calculation of
the integral in \eq{m}.\cite{Tserkovnyak:prl02} \fig{m2}
shows the calculated $|M(\kappa)|^2$.

\begin{figure}[ptb]
\includegraphics[angle=-90,width=3.375in,clip=]{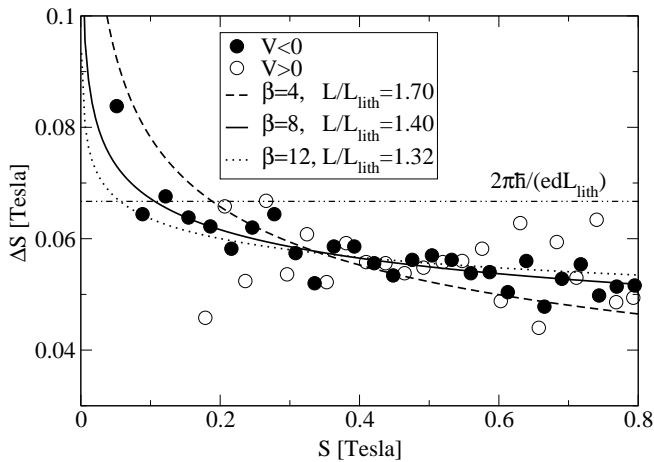}
\caption{\label{zer} Period of (faster) oscillations in
region II of \fig{osc24} as a function of
$S=\hbar\kappa_+/(ed)$. Circles show measurements at positive
and negative bias and the curves are fits using \eq{dS} at
several values of $\beta$. The best overall fit is reached
at $\beta\approx7.67$ and $L/L_{\text{lith}}\approx1.41$, where
$L_{\text{lith}}=2$~\micro{m}.}
\end{figure}

We study the profile of confinement $U_u(x)$ by measuring the
period $\Delta\kappa$ of the $|M(\kappa)|^2$
oscillations as a function of $\kappa$. In a square well of
length $L$, this period is given by $2\pi/L$. In a soft
confinement, the interference stems from the oscillations of
the electron wave function near the classical turning points,
so that $\Delta\kappa\approx2\pi/(x^+-x^-)$. For a potential of the
form
\begin{equation}
U_u(x)=E_{\text{F}u}\left|\frac{2x}{L}\right|^\beta
\label{u}
\end{equation}
(where $\beta$ characterizes the ratio between the total length of the
upper wire and the extent of its boundaries),\cite{endnote}
$2x^+/L\approx(2\kappa/k_{\text{F}})^{1/\beta}$ for
$\kappa>0$ assuming
that $\Delta k_{\text{F}}\ll k_{\text{F}}=(k_{\text{F}u}+k_{\text{F}l})/2$,
\cite{Tserkovnyak:prl02} ($x^-=-x^+$ for a symmetric potential)
and the period is therefore given by
$\Delta\kappa\approx(2\pi/L)(k_{\text{F}}/2\kappa)^{1/\beta}$.
Experimentally, we extracted $\Delta\kappa$ by measuring the distance between
oscillation zeros in region II of the interference shown in
\fig{osc24}. In order to reduce the statistical uncertainty,
the conductance was averaged along lines of constant
$\kappa_+$, separately for positive and negative bias. In
terms of variable $S=\hbar\kappa_+/ed$ (which reduces to
magnetic field $B$ at zero voltage and vanishing $\Delta
k_{\text{F}}$),
\begin{equation}
\Delta S\approx
\frac{2\pi\hbar}{edL}\left(\frac{\hbar k_{\text{F}}}{2edS}\right)^{1/\beta}\,.
\label{dS}
\end{equation}
This $\Delta S$ is compared with the data in \fig{zer} for
several values of $\beta$ (we extract
$k_{\text{F}}\approx1.5\times10^8$~m$^{-1}$ using measured
electron densities\cite{Auslaender:sc02}). At each $\beta$ shown in the
figure, the distance $L$ was found by the best (least-square) fit
of the curve (\ref{dS}) to the measurements. The lithographic
length for the junction was $L_{\text{lith}}=2$~\micro{m}
and the width $d=31$~nm.
Such fitting allows us to extract two quantities,
$L/L_{\text{lith}}=1.45\pm0.1$ and $\beta=8\pm2$,
characterizing the extent of the 1D confinement and the sharpness of
the potential-well boundaries, respectively.
It appears that the effective
length of the upper wire [defined as the distance between the
classical turning points, see \eq{u}] $L$ is actually about a
micron longer than the lithographic length.
This conclusion is relatively insensitive to the fitting procedure,
as $\Delta S$ in \eq{dS} approaches $2\pi\hbar/(edL)$ for
$S\gtrsim\hbar k_{\text{F}}/(2ed)$ if the exponent $1/\beta$ is small.
The difference between $L$ and $L_{\text{lith}}$ can
be due to significant screening of the
tungsten gates (which are positioned 0.5~\micro{m} above the junction)
by the 2DEG in the upper quantum well,
as viewed by the upper-wire electronic bands.

As a consistency check for the result of the
fit in \fig{zer}, we performed an analysis of the conductance oscillations
that takes into account the dependence on $\beta$.
According to the $S^{-1/\beta}$ scaling of the oscillations' period, see
\eq{dS}, and the $S^{1/\beta-1}$ fall-off of the their amplitude
[which follows from \eq{mpm},
also see Ref.~\onlinecite{Tserkovnyak:prl02}], the function
$S^{1-1/\beta}G\left(V,S^{1+1/\beta}\right)$ is periodic in
$S^{1+1/\beta}$. Fourier analyzing it at a fixed $V$,
and setting $\beta=8$, we obtained a main
peak, the position of which depends very weakly on $V$, which
corresponds to a length of $L=2.81\pm0.02$\micro{m}, in
agreement with the result of the fit. In \fig{osc24}b we plot
the absolute value of the main peak,
which is seen to decay on a scale of a few mV. We discuss this decay in
Sec.~\ref{dep}. For comparison, we also Fourier
analyzed the data in \fig{osc26}a. For that we
found that one has to use a larger value, $\beta=21.5$,
in order to obtain a relatively voltage-independent position of the peak.
This value of $\beta$ is reasonable because it gives approximately the same
boundary profile for
a 6~\micro{m} (upper) wire as $\beta=8$ gives for a 2~\micro{m}
wire.\cite{endnote} Again we obtained a reasonable length
($L=7.3\pm0.3$~\micro{m}) that varied only weakly as a
function of $V$. The height of the main peak in this case is
shown in \fig{osc26}b where it is seen to decay on a faster
scale than for the shorter upper wire. The ratio of the scales is
approximately the ratio of the upper-wire lengths.

\subsubsection{Modulation due to spin-charge separation}
\label{mod}

In the following we describe how electron-electron
interactions in the wires and between them affect the
oscillation pattern. We show our theoretical results for
$G(V,B)$ near the lower crossing point of the
$\left|u_1\right>\leftrightarrow\left|l_1\right>$ transition
and compare them to measurements on 2~\micro{m} and 6~\micro{m}
junctions, \figs{osc24} and \ref{osc26}. In particular, we
find that the difference in the velocities of the charge- and
spin-excitation modes in the double-wire system can account
for the observed $G(V,B)$ suppression stripes running parallel
to the $B$-axis.

As a starting point, let us consider the case when the interwire
interactions are vanishingly small $V_{ul}\ll V_{ll}$ and
the interactions in the two wires are the same,
$V_{uu}=V_{ll}$, so that $g_l=g_u=g$, as defined in \eq{gn}.
For positive voltages
$V>0$, the current (\ref{Ie}) is then given by
\begin{eqnarray}
I&=&e|\lambda|^2\int_{-\infty}^\infty dxdx^\prime\int_{-\infty}^\infty dt
e^{iq_B(x-x^\prime)}e^{ieVt/\hbar}\nonumber\\
&&\times G^>_u(x,t;x^\prime,0)G^<_l(x^\prime,0;x,t)\,.
\label{Iep}
\end{eqnarray}

At low magnetic field, the conductance has two main
contributions, corresponding to the two edge-state chiralities.
The two contributions give bright
conductance peaks and side lobes with opposite slopes, as
described in Sec.~\ref{int} and Ref.~\onlinecite{Tserkovnyak:prl02}. Let us
discuss tunneling between the right movers (current due to
tunneling between the left movers at field $B$ equals
tunneling between the right movers at field $-B$).
We assume that the electron density in each wire varies
slowly on the length scale set by the respective $k_{\text{F}}$
(except for unimportant regions very close to the boundaries).
The zero-temperature Green functions entering \eq{Iep}, in this
regime, can be written as\cite{Solyom:ap79,Meden:prb92,Voit:prb93}
\begin{eqnarray}
G_{u,l}(x,t;x^\prime,0)&=&\pm\frac{1}{2\pi}\Phi_{u,l}(x,x^\prime)\frac{1}
{(z-v_{\text{F}}t\pm i0^{+})^{\frac{1}{2}}}\nonumber\\
&&\times\frac{1}{(z-v_ct\pm i0^{+})^{\frac{1}{2}}}\nonumber\\
&&\times\left[\frac{r_c^2}{z^2-(v_ct\mp ir_c)^2}\right]
^{\frac{1}{2}\gamma}\,,
\label{gul}
\end{eqnarray}
where $v_c=v_{\text{F}}/g$, $\gamma=(g+g^{-1}-2)/4$,
$z=x-x^\prime$, and $r_c$ is a short distance cutoff (i.e., $1/r_c$
is a momentum-transfer cutoff in the electron-electron interactions).
Here, $G_u$ is the $G^>$ Green function
(\ref{ggr}) for the upper wire and $G_l$ is the $G^<$ Green
function (\ref{gle}) for the lower wire.
The function $\Phi_\nu$ is defined by
$\Phi_\nu(x,x^\prime)=\psi_\nu(x)\psi_\nu^\ast(x^\prime)$,
in terms of the WKB wave functions $\psi_\nu(x)$ for
right-moving electrons at the Fermi energy in wire $\nu$
in a confining potential $U_\nu(x)$ which must be chosen
self-consistently to give the correct electron density.
Here we assume that $\psi_u(x)$ and $U_u(x)$ are given
by Eqs.~(\ref{WKB}) and (\ref{u}), while $\psi_l(x)=e^{ik_{\text{F}l}x}$.

Several additional approximations are implied in using \eq{gul} to calculate
the tunneling current (\ref{Iep}): (1) The weak 1D-2D scattering
is neglected, (2) The voltage is small enough so that one can
linearize the noninteracting electron dispersions about the
Fermi-points and use LL theory (i.e., we disregard the curvature),
(3) $t\ll v_cL$, so that the
discreteness of the energy levels of the upper wire due to
electron confinement within a well of length $L$ and their
reflection at the boundaries does not considerably modify the
LL Green function for an infinite wire [the confinement, however,
is manifested in the form of the wave function
$\psi_u(x)$; effects due to the discreteness are discussed
in Sec.~\ref{asb} in the regime of noninteracting electrons,
and they are believed to be small]. The last approximation breaks
down for very low voltages (and, correspondingly, long times)
in the regime of the zero-bias anomaly, which is treated
separately in Sec.~\ref{lba}.

Substituting Green functions (\ref{gul}) into integral
(\ref{Iep}), we obtain for the tunneling current
\begin{equation}
I\propto\int_{-\infty}^\infty dxdx^\prime e^{i(q_B-k_{\text{F}l})
(x-x^\prime)}\psi_u(x)\psi_u^\ast(x^\prime)h(x-x^\prime)\,,
\label{Ih}
\end{equation}
using the definition
\begin{eqnarray}
h(z)&=&-\int_{-\infty}^\infty dt
\frac{e^{ieVt/\hbar}}{(z-v_{\text{F}}t+i0^+)(z-v_ct+i0^+)}\nonumber\\
&&\times\left(\frac{r_c}{z-v_ct+ir_c}\right)^\gamma
\left(\frac{r_c}{z+v_ct-ir_c}\right)^\gamma\,.
\label{hz}
\end{eqnarray}
The integrand in \eq{hz} has a simple analytic form: it has
two first-order poles at $t=z/(v_{\text{F}}+i0^+)$ and
$t=z/(v_c+i0^+)$, and two branch cuts starting with
singularities at $t=(\pm z+ir_c)/v_c$. The contour of
integration can be deformed leaving two nonvanishing
contributions: $h(z)=h_1(z)+h_2(z)$. The first contribution,
$h_1(z)$, is due to integration around the poles:
\begin{eqnarray}
h_1(z)&=&\frac{2\pi ie^{ieVz/(\hbar v_c)}}{(v_c-v_{\text{F}})(z+i0^+)}
\left(\frac{r_c^2}{r_c^2+2izr_c}\right)^\gamma\nonumber\\
&&-\frac{2\pi ie^{ieVz/(\hbar v_{\text{F}})}}{(v_c-v_{\text{F}})(z+i0^+)}\nonumber\\
&&\times\left(\frac{r_c^2}{r_c^2+z^2[1-(v_c/v_{\text{F}})^2]+
2izr_cv_c/v_{\text{F}}}\right)^\gamma\,,\nonumber\\
\label{h1}
\end{eqnarray}
and the second contribution, $h_2(z)$, is due to integration
around the branch cuts. For $z>0$, for example,
\begin{eqnarray}
h_2(z)&=&2i\sin(\gamma\pi)e^{-eVr_c/(\hbar v_c)}\left\{\int_{z/v_c}^\infty-
\int_{-\infty}^{-z/v_c}\right\} dt\nonumber\\
&&\times\left(\frac{r_c^2}{(v_ct)^2-z^2}\right)^\gamma\nonumber\\
&&\times\frac{e^{ieVt/\hbar}}{(z-v_ct-ir_c)(z-v_{\text{F}}t-ir_cv_{\text{F}}/v_c)}\,.
\label{h2}
\end{eqnarray}
In our system we expect that\cite{Auslaender:sc02}
$g\approx0.7$, so that $\gamma\approx0.03\ll1$. Therefore, since
$r_c\sim30$~nm (the width of the wires) and
$z<L\approx2-6$~\micro{m}, the terms of the form
$(\cdots)^\gamma$ in \eq{h1} can be safely ignored
(except for the regime of extremely low voltages, which will be discussed in
Sec.~\ref{lba}).
Furthermore, $h_2(z)\ll h_1(z)$, so that we arrive at an
approximation
\begin{equation}
h(z)\approx-2\pi i\frac{e^{ieVz/(\hbar v_{\text{F}})}-e^{ieVz/(\hbar v_c)}}{(v_c-v_{\text{F}})(z+i0^+)}\,.
\label{ha}
\end{equation}
Substituting this into \eq{Ih}, we can now evaluate the
current. One notices that after making the
approximation (\ref{ha}), the current (\ref{Ih}) becomes the
same as if there were no electron-electron interactions but
different Fermi velocities in the two wires, given by
$v_{\text{F}}$ and $v_c$.
Using Eqs.~(\ref{Ih}) and (\ref{ha}), we, finally, get for
the conductance $G=\partial I/\partial V$:
\begin{equation}
G(V,B)\propto\frac{1}{v_c-v_{\text{F}}}\left[\frac{1}
{v_{\text{F}}}|M(\kappa_{\text{F}})|^2-\frac{1}{v_c}|M(\kappa_c)|^2\right]\,,
\label{gvb}
\end{equation}
where $\kappa_{\text{F},c}=q_B+\Delta
k_{\text{F}}+eV/(\hbar v_{\text{F},c})$ and $M(\kappa)$ is given by \eq{m}.

If the excitation velocities in the wires are nearly the same,
$v_{\text{F}}\approx v_c=v$, we can approximate the
conductance (\ref{gvb}) by
\begin{equation}
G\propto\frac{\partial}{\partial\eta}\eta\left|M(\eta,V)\right|^2
=\left|M(\kappa)\right|^2+V\frac{\partial}{\partial
V}\left|M(\kappa)\right|^2\,, \label{ga}
\end{equation}
where $M(\eta,V)=M(q_B+\Delta k_{\text{F}}+\eta V)$
and $\eta=e/(\hbar v)$. This
reproduces \eq{ivm}. One can refine the form of the second
term on the right-hand side of \eq{ga} (which can be much
larger than the first term, see Sec.~\ref{asb}), using
approximation (\ref{mpm}), when the difference between
velocities $v_{\text{F}}$ and $v_c$ becomes appreciable
(which is the case for $g\approx0.7$):
\begin{eqnarray}
\tilde{G}&\propto&\frac{\Theta(\kappa)}{U_u^\prime(x^+)}
\cdot\frac{\sin[eVx^+(1/v_{\text{F}}-1/v_c)/\hbar]}{1/v_{\text{F}}-1/v_c}\nonumber\\
&&\times\cos2\left[\kappa x^+-s(x^+)\right]\,.
\label{gsc}
\end{eqnarray}
Here, $\kappa$ and $x^+$ are defined using velocity
$v=2(1/v_{\text{F}}+1/v_c)^{-1}$, $\tilde{G}$ stands for the second
contribution to the conductance in \eq{ga}.
At low bias, $\tilde{G}\rightarrow0$ linearly in $V$ and the
term $G\propto|M(\kappa)|^2$ governs the conductance. This
contribution is further suppressed as $V^\alpha$ (at zero temperature) in the
zero-bias anomaly regime discussed in Sec.~\ref{lba}.

We can generalize the preceding discussion of this section to
include interactions between the wires, i.e., $V_{ul}\neq0$. Since the quantum
wires are closely spaced, the interwire interactions can be
sizable. Furthermore, because the Fermi velocities in modes
$\left|u_1\right>$ and $\left|l_1\right>$ are similar, the
excitations in the coupled wires can propagate with velocities
quite different from those in the isolated wires. When we take
$V_{ul}$ into account, the dominant part of Green function
(\ref{C}) becomes (assuming weak interactions, in the spirit of the
preceding discussion)\cite{Zulicke:prb02}
\begin{eqnarray}
C(x,x^\prime;t+i0^+)&\propto&-\frac{\Phi_u\Phi_l^\ast(x,x^\prime)}
{(z-v_{\text{F}u}t)^{\frac{1}{2}}
(z-v_{\text{F}l}t)^{\frac{1}{2}}}\nonumber\\
&&\hspace{-2cm}\times\frac{1}{(z-v_{c-}t)^{\frac{1}{2}+\theta_r}
(z-v_{c+}t)^{\frac{1}{2}-\theta_r}}\,,
\label{Cz}
\end{eqnarray}
where
\begin{equation}
v_{c\pm}\approx\frac{v_{cu}+v_{cl}}{2}\pm
\frac{\tilde{V}_{ul}(0)}{\pi\hbar}\sqrt{1+r^2}\,.
\label{vcpm}
\end{equation}
Here, $r=\pi(v_{cu}-v_{cl})/2\tilde{V}_{ul}(0)$ and
$\theta_r=1/(2\sqrt{1+r^2})$ is finite for nonvanishing
interactions between the wires, $v_{cn}$ are the
charge-excitation velocities in the isolated wires.
[Note that there appears to be a sign error in
Ref.~\onlinecite{Zulicke:prb02} in the expression for the
velocities $v_{c\pm}$ in the physical case of repulsive
interactions $\tilde{V}_{ul}(0)>0$.]

For a symmetric double-wire system,
$v_{\text{F}u}=v_{\text{F}l}=v_{\text{F}}$, $V_{uu}\equiv V_{ll}$,
and $v_{cu}=v_{cl}$, so that $r=0$ and $\theta_r=1/2$. (In
this case, $v_{c+}$ and $v_{c-}$ become the velocities of the
symmetric and antisymmetric charge excitations,
respectively.) Green function then reduces to
$C\propto-\Phi_u\Phi_l^\ast(z-v_{\text{F}}t)^{-1}(z-v_{c-}t)^{-1}$ and we
reproduce our main result of this section, \eq{gvb}, after
replacing $v_c$ with the antisymmetric charge-excitation
velocity $v_{c-}$. This is natural as tunneling in a
symmetric biwire can only excite the antisymmetric
modes at low magnetic fields.

In addition to the structure studied in Sec.~\ref{asb} for
the system of noninteracting electrons, we now show that the
electron-electron interactions in the wires lead to a
modulation of the conductance oscillations along the voltage
axis.\cite{Tserkovnyak:prl02}
This modulation suppresses the contribution $\tilde{G}$
[\eq{gsc}] to zero in stripes parallel to the field axis. The
distance between them is:
\begin{equation}
\Delta V_{\text{mod}}=\frac{\pi\hbar v_{c-}v_{\text{F}}}
{ex^+(v_{c-}-v_{\text{F}})}\,.
\label{Vs}
\end{equation}
The ratio between $\Delta V_{\text{mod}}$ and the period
\begin{equation}
\Delta V=\frac{2\pi\hbar v_{c-}v_{\text{F}}}
{ex^+(v_{c-}+v_{\text{F}})}
\label{Vf}
\end{equation}
due to the wave-function oscillations near the turning points
[compare to \eq{AB}]
\begin{equation}
\frac{\Delta V_{\text{mod}}}{\Delta V}=
\frac{1}{2}~\frac{v_{c-}+v_{\text{F}}}{v_{c-}-v_{\text{F}}}=
\frac{1}{2}~\frac{1+g_-}{1-g_-} \label{sf}
\end{equation}
can be used as an independent measure of the interaction
parameter $g_-=v_{\text{F}}/v_{c-}$. From
Figs.~\ref{osc24}a and \ref{osc26}a, we find that
\begin{equation}
g_-=0.67\pm0.07\,,
\label{g-}
\end{equation}
similarly to the value for $g_l$ obtained by
the zero-bias anomaly in Sec.~\ref{zb}. Also, from \eq{gsc}
it follows that the oscillation pattern [of the principal
term $\tilde{G}(V,B)$] gains a $\pi$ phase shift across each
suppression strip. Such phase shifts can also be seen in
experimental Figs.~\ref{osc24}a and \ref{osc26}a.

\begin{figure}[ptb]
\includegraphics[width=3.375in,clip=]{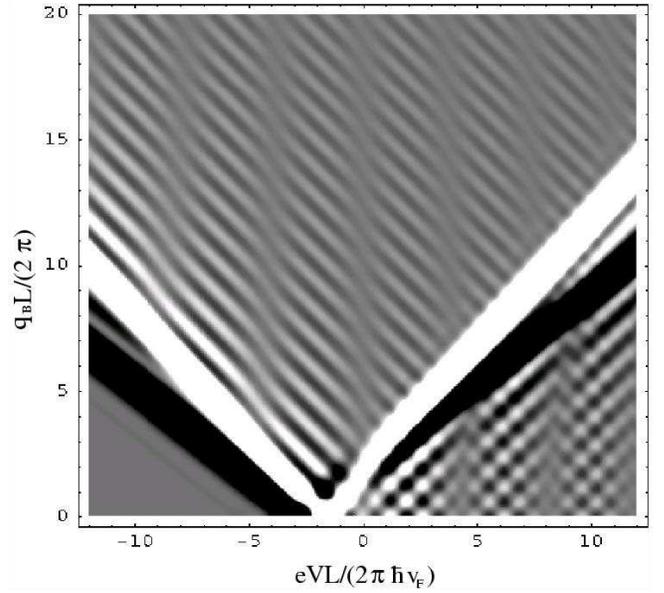}
\caption{The differential conductance interference pattern
near the lower crossing point calculated by \eq{gvb} for
tunneling between right movers (and similarly for
left movers) using a smooth confining potential for the upper
wire, \eq{u}. $v_{c-}=1.4v_{\text{F}}$, $\Delta
k_{\text{F}}=4\pi/L$, $\beta=8$. We used the numerically
found $|M(\kappa)|^2$, also shown in \fig{m2}. The figure has to be
compared to experimental \fig{osc24}.} \label{oscT}
\end{figure}

Finally, we compare the interference pattern predicted by our
theory, \eq{gvb}, with the experiment, Figs.~\ref{osc24}a,
\ref{osc26}a. $G(V,B)$ calculated using a smooth confining
potential [\eq{u} with $\beta=8$] for the upper wire is
shown in \fig{oscT}. Many pronounced features
observed experimentally--the asymmetry of the side lobes, a
slow fall-off of the oscillation amplitude and period away
from the principal peaks, an interference modulation along the
$V$-axis, $\pi$ phase shifts at the oscillation suppression
stripes running parallel to the field axis--are reproduced by
the theory.

\begin{figure}[pth]
\includegraphics[width=3.375in,clip=]{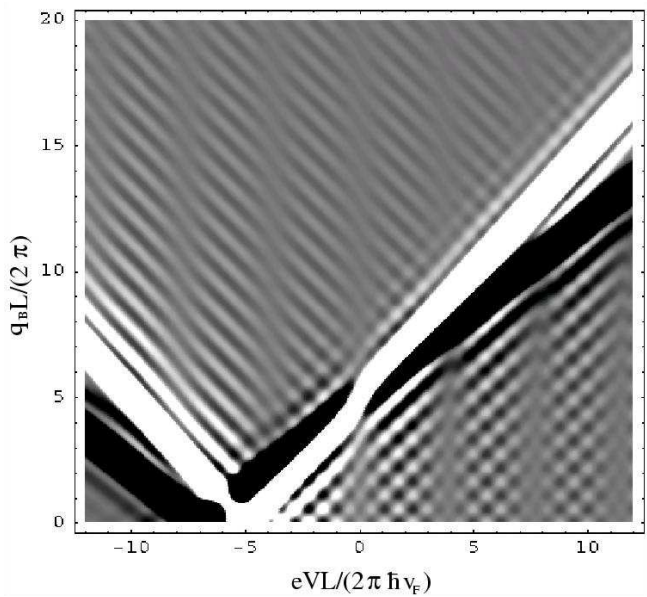}
\caption{Same as \fig{oscT} but with
$\Delta k_{\text{F}}=10\pi/L$ and $\beta=22$, describing
a longer junction with a similar boundary profile. $|M(\kappa)|^2$ was
correspondingly recomputed (now putting 300 electrons per spin in
the upper wire). The figure has to be compared to experimental \fig{osc26}.}
\label{oscT2}
\end{figure}

In \fig{oscT2}, we repeat the calculation using $\beta=22$, which defines potential (\ref{u}) with a similar boundary profile near the turning points of a three-times longer wire.\cite{endnote} (Here by length we mean the distance between the classical turning points, which, as explained in Sec.~\ref{asb}, can be somewhat different from the lithographic length.) Again an agreement between the predicted (\fig{oscT2}) and measured (\fig{osc26}) oscillation patterns is apparent. In \fig{oscT2} a few weak side lobes also appear to the left of the main dispersion peaks, unlike in \fig{oscT} where they appear strictly to the right. In addition, the interference modulation in the voltage direction has sharper features in \fig{oscT2}. These trends are expected for longer junctions as the boundaries become steeper on the scale set by the total length.

Tunneling between 1D channels with different Fermi velocities
can also yield an interference modulation similar to that
described in this section even when the electron-electron
interactions are vanishingly small.\cite{Kawamura:prb94}
It is thus important to
emphasize that we suggest the spin-charge separation picture
to explain this modulation relying on the experimental result
(see Ref.~\onlinecite{Auslaender:sc02}) that the densities of modes
$\left|u_1\right>$, $\left|l_1\right>$ and, therefore, the
corresponding Fermi velocities are nearly identical.

Using \eq{Cz} we also studied various possible scenarios
when the interactions in the two wires differ. For example,
in a situation when the upper wire is perfectly screened, so
that $V_{uu},V_{ul}\equiv0$, there are still two velocities present
in the system, $v_{\text{F}}$ and $v_{cl}$, but the
interference pattern is qualitatively very different from
that shown in \fig{oscT} and observed experimentally
[see Figs.~\ref{osc24} and \ref{osc26}]. Since a
considerable weight of the charge-excitation contribution to
the tunneling strength is shifted to velocity $v_{\text{F}}$
(which is now also the charge-excitation velocity in the
upper wire), the oscillation pattern does not exhibit the
pronounced vertical suppression stripes, but rather a much weaker
modulation. The same conclusion also holds for intermediate
regimes of relative screening in the two wires,
when the system is not symmetric and the two
charge-excitation velocities significantly differ. The
pronounced suppression stripes are, therefore, present only if
most of the charge-excitation tunneling weight is peaked at a
single velocity $v_{c-}$ (which is guaranteed only when the
system is nearly symmetric).

Taking into account 1D-2D scattering in the upper quantum wire
will smear out the oscillation pattern by its convolution with a
Lorentzian in the $B$-direction, similarly to \eq{q1} below.
The corresponding effect is, however, small because of the
high quality of our wires, which have a long scattering
length\cite{Picciotto:prl00} $l_{\text{1D-2D}}\approx6$~\micro{m}.

\subsubsection{Upper crossing point}
\label{ucp}

In practice, since the fields necessary to reach the upper
crossing point are quite large (e.g., 7~T for the
$\left|u_1\right>\leftrightarrow\left|l_1\right>$
transition), even atomic-scale disorder in the junction can
lead to a significant variation $\delta q_B$ of the momentum transfer along
the tunneling region. In particular, $\delta q_B=eB\delta d$
can be comparable with $2\pi/L$, the reciprocal wave vector of
the upper wire. This can significantly broaden the principal dispersion peaks.
Furthermore, Zeeman splitting becomes about a per cent of the Fermi energy at
these high fields and results in somewhat different dispersions for different
spin modes. Away from the main peaks, however, we still expect to see
side lobes due to stationary phases at the ends of the junction, similarly to
the regime of low magnetic fields discussed above
(with possibly a faster decoherence in the $V$
direction than just due to the dispersion curvature studied in
Sec.~\ref{dep}). Such oscillations [with about the period (\ref{AB})]
are indeed observed experimentally,
as can be seen in \fig{cross}. Because of the mentioned complications, we,
nevertheless, do not pursue a detailed analysis of the conductance near
the upper crossing point in this paper.

\subsubsection{Dephasing of the oscillations}
\label{dep}

It is evident from \figs{osc24}a and \ref{osc26}a that the
interference decays as $|V|$ is increased. A more
quantitative analysis of this decay is shown in \figs{osc24}b
and \ref{osc26}b, where the amplitude of the oscillations is
plotted as a function of voltage.
It is clear that the measured modulation has a fast-decaying envelope,
which can not be explained by the analysis of section~\ref{mod}.
(See, for example, \eq{gsc} which predicts that the modulation is
roughly periodic.)

One scenario for the dephasing occurs even in the case of
\textit{noninteracting} electrons considered in Sec.~\ref{asb},
when we take the finite curvature of the single-particle dispersions
into account. Let us return to the form of the current in \eq{I2b}:
\begin{equation}
I\propto\int_0^{eV}d\epsilon\left[|M(\kappa_+)|^2+|M(\kappa_-)|^2\right]\,.
\label{I3b}
\end{equation}
Correcting our previous results to take into account the
nonlinear dispersion near the Fermi points, we now write
$\kappa_\pm=[k^2_{\text{F}u}+2m\epsilon/\hbar^2]^{1/2}-
[k^2_{\text{F}l}+2m(\epsilon-eV)/\hbar^2]^{1/2}\pm q_B$.
[Using \eq{I3b} we still imply low enough bias $V$, so that
the density of states in the wires are relatively constant
on the energy scale of $e|V|$.]
Expanding this expression
to lowest order in curvature, we further obtain
\begin{equation}
\kappa_\pm=\Delta k_{\text{F}}+\frac{eV}{\hbar
v_{\text{F}}}\pm q_B+\frac{eV(eV-2\epsilon
)}{2\hbar^2v^2_{\text{F}}k_{\text{F}}}\,.
\label{kappaexp}
\end{equation}
[\eq{ivm} can be recovered by neglecting the last term above.]
The current (\ref{I3b}) then becomes
\begin{eqnarray}
I&\propto&\int_{-eV/2}^{eV/2}d\epsilon\left|M\left(\Delta k_{\text{F}}+
\frac{eV}{\hbar v_{\text{F}}}+q_B-\frac{\epsilon
eV}{\hbar^2v^2_{\text{F}}k_{\text{F}}}\right)\right|^2\nonumber\\
&+&(q_B\rightarrow-q_B)\,.
\label{I4b}
\end{eqnarray}

It is easy to see now that the contribution to the
conductance obtained by differentiating the integrand in
\eq{I4b} will be suppressed when the argument $\kappa$ of the
tunneling matrix amplitude $M(\kappa)$ changes by the full
period of oscillations $\Delta\kappa$ upon energy $\epsilon$
variation between the integration limits $\pm
eV_{\text{sup}}/2$. We thus arrive at the condition for the
suppression voltage $V_{\text{sup}}$:
\begin{equation}
\Delta\kappa=\frac{(eV_{\text{sup}})^2}{\hbar^2v^2_{\text{F}}k_{\text{F}}}\,.
\end{equation}
Approximating $\Delta\kappa\approx2\pi/L$ and translating it
into the oscillation period in the bias direction $e\Delta
V=\hbar v_{\text{F}}\Delta\kappa$, one finally obtains
\begin{equation}\label{Vsup}
\frac{V_{\text{sup}}}{\Delta
V}=\sqrt{\frac{Lk_{\text{F}}}{2\pi}}\,.
\end{equation}
Using density 100~\micro{m}$^{-1}$ for the lowest bands in the
wires,\cite{Auslaender:sc02} we find
$V_{\text{sup}}/\Delta V\approx7$ ($\approx12$) for the
2~\micro{m} (6~\micro{m}) junction. An implicit assumption in
the derivation is that we are still close enough to the Fermi
level so that higher-order corrections should not modify the
result significantly [in particular, for the calculation of
the matrix element (\ref{m}) it can still be reasonable to use
the wave function $\psi_u(x)$ at the Fermi energy].

\begin{figure}[ptb]
\includegraphics[width=3.375in,angle=0,clip=]{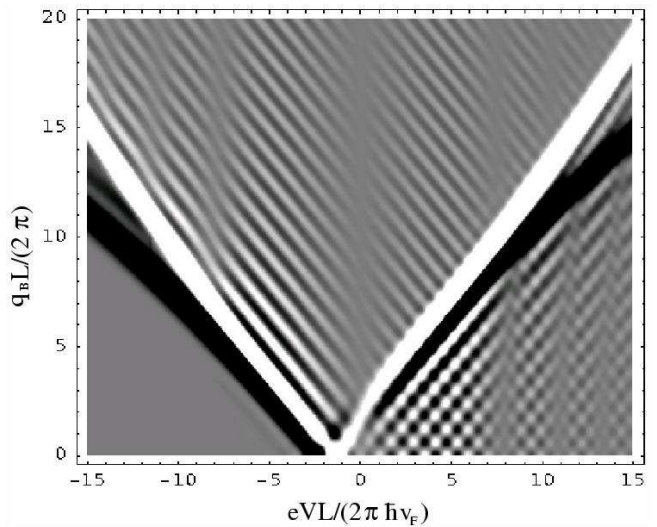}
\caption{\label{dephasing}The differential conductance
interference pattern near the lower crossing point calculated
by \eq{I4b}, within the noninteracting electron picture,
using the matrix element $M(\kappa)$ shown in \fig{m2}
(for $\beta=8$). See text for further details.}
\end{figure}

The result of the numerical calculation using \eq{I4b} and the
matrix element $M(\kappa)$ plotted in \fig{m2} (using parameters
characteristic for the 2~\micro{m} sample) is shown in \fig{dephasing}.
Notice that when the voltage exceeds $V_{\text{sup}}\approx7\Delta V$,
so that the pattern starts dephasing due to the finite curvature,
a beating pattern appears.
It differs from the data in several important aspects: First of all,
the lines of suppressed \GVB\ are not equidistant.
In addition, $V_{\text{sup}}$, corresponding to the first
suppression stripe (on either the positive- or negative-voltage sides),
is about twice larger than the period we observe in
\fig{osc24}b and four times larger than that in \fig{osc26}b,
which in both cases is given by about $3\Delta V$. This suggests that the
source of the beating in the experimental data is not the curvature of the
dispersions, but rather the spin-charge separation mechanism discussed
in Sec.~\ref{mod}.

Another important difference between
\eq{I3b} and the experiment is that the decay of the
oscillations is much stronger in the latter. It might therefore be necessary to
consider both the curvature and electron interactions in order
to understand the fast decay of the conductance oscillation
amplitude with increasing voltage. Taking into account
the curvature while bosonizing excitations of the
\textit{interacting} electrons\cite{Haldane:jpc81,Voit:rpp94}
leads to higher-order terms in the Hamiltonian.
Physically this corresponds to interactions between bosonic excitations
which therefore acquire a finite life time. The singularities of
the spectral densities will correspondingly be rounded, in turn smearing
the conductance interference pattern.
Further complications may arise from the
electron backscattering which was entirely disregarded: While the
low-energy properties of the system are not affected by the
backscattering (apart from rescaling of certain parameters)
since it renormalizes downward in the case of repulsive interactions,
the story at a finite energy could be different. The reason for this
is a slow (logarithmic) renormalization flow of the backscattering
strength. If a significant backscattering is present in the original
Hamiltonian, it could therefore be still considerable at a finite energy.
A detailed study of these effects however lies beyond this paper's scope.

\subsection{Zero-bias anomaly}
\label{zba}

\subsubsection{Crossing points}
\label{lba}

It is enlightening to further study tunneling between 1D channels at
low bias when the magnetic field is tuned to match two
Fermi-points of the wires (see Sec.~\ref{disp}). The
zero-bias properties are similar near the two crossing points
and, for definiteness, we choose to discuss the upper
crossing, where the magnetic wave vector $q_B$ is close to
$k_{\text{F}u}+k_{\text{F}l}$ and the field changes the chirality
of the tunneling electrons: The tunneling is amongst the left movers of the upper wire and the right movers of the lower wire.
For the $\left|u_1\right>\leftrightarrow\left|l_1\right>$ transition,
this point is located at $B\approx7$~T, see \fig{GVB}. The
results are straightforward to apply to the regime of
the lower crossing point, as well.

For clarity, we start by making a series of simplifying
assumptions which will be dropped in subsequent generalizations:
First, we set the upper-wire and interwire interactions, $V_{uu}$
and $V_{ul}$, to zero. Physically, this corresponds to a
regime where the Coulomb interactions in the upper wire are
perfectly screened by the 2DEG. Secondly, we further simplify the
model by assuming a square-well confinement for the electronic
states in the upper quantum wire and an infinitely-steep
reflecting left boundary for the electrons in the lower wire,
i.e., $U_u(x)$ [$U_l(x)$] is constant for $|x|<L/2$
[$x>-L/2$] and infinite otherwise. As we showed in the previous
sections, both of the above assumptions are not very realistic for
the purpose of studying the interference pattern. In the zero-bias
anomaly regime, however, they can be a good starting
point, at least, for pedagogical reasons.

Electron states participating in tunneling near the
crossing points [\eq{B12}] lie close to the Fermi levels in
both wires. It is therefore possible to calculate the
correlation functions analytically using LL theory, after the
dispersion relations in the wires are linearized. At the
upper crossing point, we only need to retain Green
functions of the left movers of the upper wire and the
right movers of the lower wire. At zero temperature these
are given by
\begin{eqnarray}
G^>_u(x,t+i0^+;x^\prime,0)&=&-\frac{1}{4L}\frac{e^{-ik_{\text{F}u}z}
e^{-\Gamma|z|/v_{\text{F}}}}{\sin\frac{\pi}{2L}(z+v_{\text{F}}t)}\nonumber\\
&\stackrel{L\rightarrow\infty}{=}&-\frac{1}{2\pi}\frac{e^{-ik_{\text{F}u}z}
e^{-\Gamma|z|/v_{\text{F}}}}{z+v_{\text{F}}t}
\label{Gu}
\end{eqnarray}
for $|x|,~|x^\prime|<L/2$, and $G^>_u$ vanishing otherwise, and
\begin{eqnarray}
G^<_l(x^\prime,0;x,t+i0^+)&=&-\frac{1}{2\pi}
\frac{e^{-ik_{\text{F}l}z}}{(z-v_{\text{F}}t)^{\frac{1}{2}}}\nonumber\\
&&\hspace{-1cm}\times\frac{1}{(z-v_{cl}t)^{\frac{1}{2}}}\nonumber\\
&&\hspace{-1cm}\times\left[\frac{r_c^2}{z^2-(v_{cl}t-ir_c)^2}\right]
^{\frac{g_l+g_l^{-1}-2}{8}}\nonumber\\
&&\hspace{-1cm}\times\left[\frac{z^{\prime2}-z^2}
{z^{\prime2}-(v_{cl}t)^2}\right]^{\frac{g_l-g_l^{-1}}{8}}\,,
\label{Gl}
\end{eqnarray}
for $x,x^\prime>-L/2$, and vanishing otherwise, where
$z=x-x^\prime$, $z^\prime=x+x^\prime+L$, and $r_c$ is a small
distance cutoff. As specified above, \eq{Gu} [\eq{Gl}]
contains only the component for the left (right) movers in
the upper (lower) wire; we have thus omitted terms
proportional to $e^{ik_{\text{F}}z}$,
$e^{ik_{\text{F}}z^\prime}$, and $e^{-ik_{\text{F}}z^\prime}$
which do not contribute constructively to tunneling near the
upper crossing point. The last factor in the expression for
$G^<_l$ is due to the closed boundary at
$x=-L/2$.\cite{Eggert:prl96,Mattsson:prb97,Fabrizio:prb95}

For sufficiently large voltages, $eV\gg2\hbar v_{\text{F}}/(g_lL)$,
the tunneling electrons do not feel the junction boundaries on
the time scale set by the voltage. In particular,
the left boundary of the lower wire does not affect
the dynamics and, effectively, electrons directly tunnel into
the bulk of the lower wire: The last term
in \eq{Gl} is close to unity and can, therefore, be omitted.
Terms of the form $1/(z\pm vt)^\vartheta$ entering
Eqs.~(\ref{Gu}) and (\ref{Gl}) are dominated by the long-$t$
behavior in the integral [\eq{Iep}, the voltage is assumed to
be positive] if $eV\ll\hbar\max(vq,\Gamma)$, where
$q=q_B-(k_{\text{F}u}+k_{\text{F}l})$. The conductance is then
suppressed as a power law
\begin{equation}
G(V)\propto V^{\alpha}
\label{GV}
\end{equation}
with the exponent
$\alpha_{\text{bulk}}=(g_l+g_l^{-1}-2)/4$. This result is
easy to generalize for the case of unscreened interactions in
the upper wire:
\begin{equation}
\alpha_{\text{bulk}}=\sum_{\nu=u,l}\frac{g_\nu+g_\nu^{-1}-2}{4}\,.
\label{ab}
\end{equation}
If the interwire interactions $V_{ul}$ are also significant,
the elementary excitation modes in the wires become coupled
and $\alpha_{\text{bulk}}$ has a more complicated form than
that in \eq{ab}.\cite{Carpentier:prb02}
Interference oscillations discussed in Sec.~\ref{int} can modulate
the power-law current suppression (\ref{GV}), setting an upper voltage bound,
$eV<e\Delta V\approx2\pi\hbar v_{\text{F}}/L$, for the validity of \eq{GV}.
It would therefore be hard to observe the exact power-law voltage dependence
(\ref{GV}) with the exponent (\ref{ab}) in the regime when
$eV\gg2\hbar v_{\text{F}}/(g_lL)$ (see, however, Sec.~\ref{zb}).

If $eV\ll2\hbar v_{\text{F}}/(g_lL)$, electrons effectively
tunnel into the end of the lower wire and the current suppression
is governed by processes in the lower wire outside the tunneling region.
In particular, details of
the interactions in the finite upper wire do not play a role.
The last term in \eq{Gl} now also contributes to the
exponent of the long-$t$ asymptotic, and $\alpha$ in \eq{GV} is
given by
\begin{equation}
\alpha_{\text{end}}=\frac{g_l^{-1}-1}{2}\,.
\label{ae}
\end{equation}
The upper wire, in this
case, can be viewed as a point contact and the tunneling
exponent is determined entirely by the properties of the lower
wire outside the tunneling region.

At a finite temperature, the time scale relevant for the discussion above
is set by $\max(eV,k_BT)$. The power law (\ref{GV}) should now
be replaced with
\begin{equation}\label{GVT}
G(V,T)\propto T^\alpha F_\alpha\left(\frac{eV}{k_BT}\right)\,,
\end{equation}
where $F_\alpha(x)$ is a known scaling function with properties
$F_\alpha(0)=\text{const}$ and $F_\alpha(x)\propto x^\alpha$ in the limit
of $x\gg1$.\cite{Bockrath:nat99}
At low temperatures the conductance yields a
low-bias dip extending to voltages $eV\sim k_BT$
with $G(V=0)\propto T^\alpha$.

In Sec.~\ref{int} we showed that the conductance $G(V,B)$
exhibits a characteristic interference pattern due to
wave-function oscillations near the gates confining the
tunneling region. We can easily read out the profile of this
pattern for the current (\ref{Iep}) using the correlation
functions (\ref{Gu}), (\ref{Gl}) in the low-energy
regime considered in this section (namely $t\gg z$):
\begin{equation}
G(B)\propto\int_{-\infty}^{\infty}dk\frac{\Gamma/v_{\text{F}}}
{k^2+(\Gamma/v_{\text{F}})^2}|M(k-q)|^2\,,
\label{q1}
\end{equation}
where $M(\kappa)$ is the tunneling matrix element, \eq{m}.

The discussion in this section also holds for the lower
crossing point, where the electrons do not change their
chirality upon tunneling. To directly apply the above results
to this regime (for definiteness, assuming we now consider
the transition between the right-moving electrons), we only
need to redefine the distance from the crossing point in the
field direction: $q=q_B+k_{\text{F}u}-k_{\text{F}l}$ (and
analogously for the transition between the left movers).

\subsubsection{Direct tunneling from the 2DEG}
\label{zb}

It is straightforward to generalize the main results
of the preceding section to the regime of
direct tunneling from the 2DEG. \eq{Iep} stays valid in
this case, but now $G_u^>$ is Green function for the 2DEG
near the edge of the upper quantum well. We calculate this
correlation function and discuss its limiting behavior at low
energies in Appendix~\ref{a2}. The 2DEG density of states is
finite at the Fermi energy and, therefore, the long-$t$
behavior of the one-particle Green function is
$G^>(t)\propto1/t$. If $\max(eV,k_BT)\ll\hbar
v_{\text{F}}k_{\text{F,2D}}$, where $\hbar k_{\text{F,2D}}$
is the 2DEG Fermi momentum and $v_{\text{F}}$ is the lower of the Fermi
velocities of the 1D band and the 2DEG, the temperature and voltage
dependence of the differential conductance are governed by
the exponents (\ref{ab}), with $g_u=0$, or (\ref{ae}), depending
on the relation between $\max(eV,k_BT)$ and $2\hbar
v_{\text{F}}/(g_lL)$. Because in this regime we tunnel directly
from the 2DEG, interactions in the 1D modes of the upper
quantum well do not play a role, and both
$\alpha_{\text{bulk}}$ and $\alpha_{\text{end}}$ are determined
only by the interaction constant $g_l$ of the lower wire.
While the field dependence of the conductance for the direct
2DEG--lower wire tunneling is different from \eq{q1} (in
particular, the conductance does not exhibit a strong
oscillation pattern), the low-energy properties stay similar
to the case of the 1D-1D tunneling.
In spite of a complicated dependence of $G(V,B)$ on
magnetic field, the zero-bias anomaly is pronounced in the data for
tunneling either between different 1D bands or between the 2DEG and
the 1D bands.

As described in Sec.~\ref{dtc}, we measured the zero-voltage
conductance dip at temperatures $0.2<T<2$~K on a junction of
length $L=6$~\micro{m} at $B=2.5$~T. It can be seen in
\fig{GVB} that at this magnetic field, the conductance is
dominated by direct tunneling from the 2DEG,
$\left|u_3\right>\leftrightarrow\left|l_{2}\right>$.
Since $\hbar
v_{\text{F}}k_{\text{F,2D}}/k_B\sim100$~K$\gg T$, the
temperature dependence of the zero-bias dip can be used to
extract the value of the interaction constant $g_l$ for the
band $\left|l_2\right>$. The data points and the (best)
theoretical fitting curves are shown in
\fig{zbd}; we find
\begin{equation}
g_l=0.59\pm0.03\,.
\label{gl}
\end{equation}
The transition point between the two lines
in the plot is consistent with an estimate $2\hbar
v_{\text{F}}/(g_lLk_B)\approx0.5$~K for the second 1D
mode of the lower wire, $\left|l_2\right>$.

As a consistency check, we plot in the insets to \fig{zbd}
curves calculated using \eq{GVT} (taking
both $\alpha_{\text end}$ and $\alpha_{\text bulk}$
for the exponent). $g_l$ and the overall
proportionality constants were independently obtained from the power-law
temperature dependence of the bottom of the dip, i.e., $G(V=0,T)$,
so that at this point we do not have any remaining fitting parameters.
The results show reasonable agreement with the data: When
${\text{max}}\left(eV,k_BT\right)>2\hbar v_{\text{F}}/(g_lL)$
the data is consistent with $\alpha=\alpha_{\text bulk}$
while when ${\text{max}}\left(eV,k_BT\right)<2\hbar
v_{\text{F}}/(g_lL)$ it is more consistent with
$\alpha=\alpha_{\text end}$. Thus, in particular, there is a crossover
between $\alpha_{\text end}$ and $\alpha_{\text bulk}$
in the data for $G(V)$ at $T=0.24$~K. For voltages $V\sim1$~meV that are
comparable to the Fermi energies of the modes participating
in tunneling, the power-law behavior (\ref{GVT}) is replaced by
a more complex structure modulated by the dispersions in
the wires and the upper well, see \fig{GVB}.

\section{Conclusions}

We have presented a detailed experimental and theoretical
investigation of tunneling between two interacting quantum
wires of exceptional quality fabricated at the cleaved edge of
a GaAs/AlGaAs heterostructure. The study focused on
revealing electron-electron interaction effects on the
conductance interference pattern arising from the finite size of
the tunneling region and the conductance suppression at low
voltage.

In the analysis of the data the finiteness of the junction
plays a central role. Breaking translational invariance, the
boundaries give rise to secondary dispersion peaks in the
dependence of the conductance on voltage bias and magnetic
field. Smooth gate potentials result in a strongly asymmetric
interference profile, while the Coulomb repulsion in the
wires leads to spin-charge separation which, in turn,
modulates the conductance oscillation amplitude
as a function of voltage bias.

An interplay between the electron correlations in the
wires and the finiteness of the junction length also results in
different regimes of the zero-bias anomaly. At the lowest
voltages, the upper wire is effectively a point-contact
source for injecting electrons into the semi-infinite lower
wire. On the other hand, at higher voltages, electrons
effectively tunnel between the bulks of the two wires along
the length of the junction.

Using the temperature dependence of the zero-bias dip, we
found the value of the interaction parameter
$g_l=v_{\text{F}l}/v_{cl}$ for band
$\left|l_2\right>$ in the lower wire to be
$0.59\pm0.03$. From the ratio between the long (due to
spin-charge separation) and slow (due to upper-wire
confinement) scales of the conductance oscillations, we also
extracted the interaction parameter $g_-=v_{\text{F}}/v_{c-}$
corresponding to the antisymmetric charge-excitation mode in
the lowest bands $\left|u_1\right>$ and $\left|l_1\right>$ of
the biwire to be $0.67\pm0.07$.

While $g_-$ and $g_l$ have similar numerical values,
these quantities should be contrasted:
$g_l$ is the interaction parameter (\ref{gn}) of the
channel $\left|l_2\right>$ in the lower wire, which is screened
by other 1D states in the wires as well as the 2DEG of the upper quantum well.
$g_-$, on the other hand, is a parameter characterizing the (antisymmetric)
charge mode in the coupled $\left|u_1\right>$ and $\left|l_1\right>$
channels of the two wires, which is relatively weakly screened by the
2DEG since the latter has a smaller Fermi velocity (being, nevertheless,
still larger than the Fermi velocity of
$\left|l_2\right>$).\cite{Auslaender:sc02}
This can explain why $g_-$ and $g_l$ are comparable while
$\left|l_2\right>$ has about half the Fermi velocity of
$\left|u_1\right>$ and $\left|l_1\right>$. [The interwire
interaction would only enhance the mismatch as it reduces $v_{c-}$,
see \eq{vcpm}].]

Similar values for the interaction parameter $g$, in the range
between 0.66 and 0.82,
were found in Ref.~\onlinecite{Auslaender:prl00} for single
cleaved-edge quantum wires by measuring the temperature dependence
of the line width of resonant tunneling through a localized impurity state.
Spectral properties of the same double-wire structure as reported
here were investigated in Ref.~\onlinecite{Auslaender:sc02},
also indicating comparable values of $g$, about 0.75, for various intermode
transitions. An interaction parameter $g\approx0.4$ was found for GaAs
quantum-wire stacks in resonant Raman scattering experiments;\cite{Rubio}
the smaller value of $g$ there can be attributed to much lower electron
densities and no screening by the 2DEG as in our measurements.

We have enjoyed illuminating discussions with Y. Oreg and A.
Stern. This work was supported in part by the US-Israel BSF,
NSF Grant DMR 02-33773, and by a research grant from the
Fusfeld Research Fund. OMA is supported by a grant from the
Israeli Ministry of Science.
\appendix

\section{Independent-mode approximation}
\label{a1}

In our analysis we treat different 1D bands in the wires as
independent and disregard interband interactions. While
this is a convenient approximation for theoretical
investigations that has been often assumed in previous
works,\cite{Auslaender:sc02,Tserkovnyak:prl02,Carpentier:prb02,Zulicke:prb02}
it needs to be further justified. Tunneling into multimode 1D
wires was considered in Ref.~\onlinecite{Matveev:prl93}.
It was shown that low-energy tunneling into the edge of a
semi-infinite wire with $N$ bands is governed by the
tunneling density of states exponents
$\alpha_i$ such that the differential conductance (at zero temperature and low voltage $V$) is given by
$G\propto\sum_{i=1}^N|t_i|^2V^{\alpha_i}$, where $t_i$ is the
tunneling amplitude for the $i$th mode. In the independent-mode
approximation with interactions described by Hamiltonian
(\ref{Hint}) for each mode, these exponents are given by
\eq{ae} with the parameter $g$ describing interactions in
each mode. On the other hand, in a more realistic picture one
deals with an interaction Hamiltonian
\begin{equation}
H_{\text{int}}=\frac{V_0}{2}\sum_{i,j=1}^N\int_0^\infty dx\rho_i(x)\rho_j(x)
\end{equation}
which takes into account the interband coupling. Here, $V_0$
is the zero-momentum Fourier component of the interaction
potential $V(x)=V_0\delta(x)$ and $\rho_i$ is the electron
density in the $i$th band. The exact form of the potential is not
important as we are only interested in the long--wave-length
quantum fluctuations.\cite{Matveev:prl93}

The exponents are given by\cite{Matveev:prl93}
$\alpha_i=(\sum_{l=1}^N\gamma^2_{il}s_l/v_i)-1$, where $v_i$
is the Fermi velocity of the noninteracting 1D electron gas
at the density of the $i$th mode, $s_l$ is the velocity of the
$l$th soundlike excitation in the presence of the
potential $V(x)$, and $\gamma_{il}$ characterizes coupling
between the $i$th and $l$th noninteracting modes after the
interaction potential $V(x)$ is switched on. In the case of a
single transverse mode with spin degeneracy, $N=2$,
$\gamma^2_{il}=1/2$, and $s_1=v_{\text{F}}\sqrt{1+2V_0/(\pi\hbar
v_{\text{F}})}$, $s_2=v_{\text{F}}$ are the charge- and
spin-excitation velocities, respectively. For a general $N$,
the velocities $s_l$ are given by roots of the equation
\begin{equation}
\sum_{i=1}^N\frac{v_i}{s_l^2-v_i^2}=\frac{\pi\hbar}{V_0}
\end{equation}
and the coefficients $\gamma_{il}$ are given by
\begin{equation}
\gamma^2_{il}=\frac{v_i}{(s_l^2-v_i^2)^2}
\left[\sum_{j=1}^N\frac{v_j}{(s_l^2-v_j^2)^2}\right]^{-1}\,.
\end{equation}
In our system,\cite{Auslaender:sc02} the Fermi velocities of
the highest occupied bands are very different (e.g.,
the highest transverse mode has twice the velocity of the next
lower-lying mode). Furthermore, since the interaction
$V_0\lesssim\max(\hbar v_i)$ is not too large, the correction
to the exponents $\alpha_i$ due to the interband coupling is
expected to be relatively small. One can
accommodate for this correction by slightly renormalizing the
interaction constants $g$, viewing it as a mutual interband
screening.\cite{Matveev:prl93}

Also, it is safe to disregard intermode transitions
as they are determined by the
Fourier components of the interaction with a large wave vector
$k\sim k_{\text{F}}$, which are small for a smooth long-range
potential.\cite{Matveev:prl93} The weak backscattering within
each spin-degenerate mode can be further renormalized downward at low
energies in the physical case of repulsive interactions.\cite{Solyom:ap79}

\section{Direct tunneling from the 2DEG}
\label{a2}

In order to describe the $V$ and $B$ dependence of the
conductance for direct 2DEG--lower wire tunneling, we
approximate the Green function of the top quantum well by the
edge Green function of a 2D electron gas occupying a
half plane $y>0$ with $x$ extended from $-\infty$ to $\infty$.
We assume the potential is $V(x,y)=0$ for $y>0$ and
$V(x,y)=\infty$ for $y<0$. Therefore, we find
\begin{eqnarray}
iG^>(x,y,t;x^\prime,y^\prime,0)&=&\frac{1}{\pi^2}
\int_{-\infty}^\infty dpe^{ip(x-x^\prime)}\int_0^\infty dk\nonumber\\
&&\times\sin(ky)\sin(ky^\prime)\Theta(\epsilon)e^{-i\epsilon t/\hbar}\,,\nonumber\\
\end{eqnarray}
where $\epsilon=\hbar^2(p^2+k^2-k_{\text{F}}^2)/(2m)$ is the
energy and $k_{\text{F}}^2$ is the Fermi wave vector of the
2DEG. $\Theta(\epsilon)$ is the Heaviside step function. When
we calculate the tunneling current, $y$ and $y^\prime$ run
from $0$ to $\xi$, the width of the tunnel junction (i.e.,
the extent of the 1D mode of the lower wire in the
direction perpendicular to the cleaved edge). We set
$(y,y^\prime)\rightarrow\xi/2$ and approximate
$\sin(k\xi/2)\approx k\xi/2$ assuming $k_{\text{F}}<1/\xi$.
In the frequency domain, Green function
$G^>(z,\omega)=\int_{-\infty}^\infty dte^{i\omega
t}G^>(z,t)$, with $z=x-x^\prime$, then becomes
\begin{equation}
iG^>(z,\omega)=\frac{\xi^2}{2\pi\hbar}\int_{-\infty}^\infty dpe^{ipz}
\int_0^\infty dkk^2\delta(\epsilon-\omega)\Theta(\omega)\,.
\end{equation}
In the limit of small positive frequencies it reduces to
\begin{eqnarray}
iG^>(z,\omega\rightarrow0^+)&=&m\frac{\xi^2}{2\pi\hbar}
\int_{-k_{\text{F}}}^{k_{\text{F}}}dpe^{ipz}\sqrt{k_{\text{F}}^2-p^2}\nonumber\\
&=&m(\xi k_{\text{F}})^2\frac{J_1(k_{\text{F}}z)}{2\hbar k_{\text{F}}z}\,,
\label{G2D}
\end{eqnarray}
where $J_1$ is the first-order Bessel function of the first
kind. In particular, since $J_1(x)\propto x$ when
$x\rightarrow0$, the density of states is finite at the Fermi
energy and $G^>(t)\propto1/t$ as $t\rightarrow\infty$.
Furthermore, from the low-energy form of the 2DEG Green
function [\eq{G2D}] it follows that the relevant range of $z$
in integral (\ref{Ie}) is $1/k_{\text{F}}$ rather than
$1/\max(q,\Gamma/v_{\text{F}})$ as in the case of the 1D-1D tunneling.


\begin{thebibliography}{20}
\bibitem{endnote} If we assume $\beta=L/l$ in \eq{u}, where $l$ is a fixed
length, then for large $L$, $U_u(x)$ takes the form which is independent of
$L$ at the boundaries: $U_u(x)\approx E_{\text{F}u}\exp[-2(L/2-|x|)/l]$. Note:
$\alpha$ was used in Ref.~\onlinecite{Tserkovnyak:prl02} instead of $\beta$
in \eq{u}; in this paper $\alpha$ denotes the LL exponent, see, e.g., \eq{GV}.
\end{thebibliography}
\end{document}